\documentclass[modern,trackchanges]{aastex62}

\usepackage{bm} 
\usepackage{amsmath}
\usepackage{multirow}

\newcommand*\diff{\mathop{}\!\mathrm{d}}
\newcommand{\Secref}[1]{\hyperref[#1]{Section~\ref*{#1}}}
\newcommand{\Appref}[1]{\hyperref[#1]{Appendix~\ref*{#1}}}

\graphicspath{{./}{figures/}}

\submitjournal{AJ}

\shorttitle{RV Measurements of HR 8799 b and c}
\shortauthors{Ruffio et al.}

\begin{document}

\title{Radial Velocity Measurements of HR 8799 b and c with Medium Resolution Spectroscopy}

\correspondingauthor{Jean-Baptiste Ruffio}
\email{jruffio@stanford.edu}

\author[0000-0003-2233-4821]{Jean-Baptiste Ruffio}
\affiliation{Kavli Institute for Particle Astrophysics and Cosmology, Stanford University, Stanford, CA 94305, USA}

\author[0000-0003-1212-7538]{Bruce Macintosh}
\affiliation{Kavli Institute for Particle Astrophysics and Cosmology, Stanford University, Stanford, CA 94305, USA}

\author[0000-0002-9936-6285]{Quinn M. Konopacky}
\affiliation{Center for Astrophysics and Space Science, University of California, San Diego; La Jolla, CA 92093, USA}

\author[0000-0002-7129-3002]{Travis Barman}
\affiliation{Lunar and Planetary Laboratory, University of Arizona, Tucson, AZ 85721, USA}

\author[0000-0002-4918-0247]{Robert J. De Rosa}
\affiliation{Kavli Institute for Particle Astrophysics and Cosmology, Stanford University, Stanford, CA 94305, USA}

\author[0000-0003-0774-6502]{Jason J. Wang}
\altaffiliation{51 Pegasi b Fellow}
\affiliation{Department of Astronomy, California Institute of Technology, Pasadena, CA 91125, USA}

\author[0000-0002-9803-8255]{Kielan K. Wilcomb}
\affiliation{Center for Astrophysics and Space Science, University of California, San Diego; La Jolla, CA 92093, USA}

\author[0000-0002-1483-8811]{Ian Czekala}
\altaffiliation{NASA Hubble Fellowship Program Sagan Fellow}
\affiliation{Astronomy Department, University of California, Berkeley; Berkeley, CA 94720, USA}

\author[0000-0002-4164-4182]{Christian Marois}
\affiliation{National Research Council of Canada Herzberg, 5071 West Saanich Rd, Victoria, BC, Canada V9E 2E7}
\affiliation{University of Victoria, 3800 Finnerty Rd, Victoria, BC, Canada V8P 5C2}

\begin{abstract}
High-contrast medium resolution spectroscopy has been used to detect molecules such as water and carbon monoxide in the atmospheres of gas giant exoplanets. In this work, we show how it can be used to derive radial velocity (RV) measurements of directly imaged exoplanets.
Improving upon the traditional cross-correlation technique, we develop a new likelihood based on joint forward modelling of the planetary signal and the starlight background (i.e., speckles).
After marginalizing over the starlight model, we infer the barycentric RV of HR 8799 b and c in 2010 yielding $-9.2 \pm 0.5\,\mathrm{km/s}$ and $-11.6 \pm 0.5\,\mathrm{km/s}$ respectively. 
These RV measurements help to constrain the 3D orientation of the orbit of the planet by resolving the degeneracy in the longitude of ascending node. Assuming coplanar orbits for HR 8799 b and c, but not including d and e, we estimate $\Omega={89^{\circ}}^{\,+27}_{\,-17}$ and $i={20.8^{\circ}}^{\,+4.5}_{\,-3.7}$. 
\end{abstract}


\section{Introduction}

The four planets orbiting the star HR 8799 have been poster children of the direct detection of exoplanets since their discovery \citep{Marois2008science,Marois2010}. They orbit around their star at semi-major axes between $15-70\,\mathrm{au}$ for a system $\approx41\,\mathrm{pc}$ \citep{Gaia2018} away from the Sun. Their masses lie between $5-10\, M_{\mathrm{Jup}}$, which are derived from their luminosity, the estimated age of the system ($\approx 41\,\mathrm{Myr}$; \citealt{Bell2015,Zuckerman2011}) and evolutionary models \citep{Baraffe2003}. The planets are surrounded by warm ($<10\,\mathrm{au}$) and cold ($>100\,\mathrm{au}$) dust belts, similar to the asteroid and the Kuiper belt in our own solar system \citep{Su2009,Reidemeister2009,Hughes2011,Matthews2014,Booth2016}.

The orbital parameters of the four planets have been constrained by a decade's worth of monitoring \citep{Fabrycky2010,Soummer2011,Pueyo2015,Konopacky2016,Zurlo2016, Wertz2017, Wang2018, ONeil2019}. The planets are thought to only slightly deviate from co-planarity, with a small inclination around $27^{\circ}$, and close to 1:2:4:8 resonance \citep{Wang2018}. In a recent milestone, \citet{Lacour2019} used optical interferometry to measure $100\,\mu\mathrm{as}$ astrometry of the closest planet in the system (HR 8799 e). Unfortunately, direct imaging data does not distinguish between in-the-plane or out-the-plane of the sky motion of the planets, leading in particular to a $180^{\circ}$ degeneracy in the longitude of the ascending node.
Measuring the radial velocity (RV) of the planets can provide important constraints to the 3D orientation of the orbits, but this requires medium ($R=\lambda/\diff\lambda\sim4000$) to high ($R>25,000$) resolution spectra (HRS).
HR 8799 b and c have already been studied at medium resolution with Keck/OSIRIS ($R\approx4000$; \citealt{Larkin2006}), for example providing unambiguous detection of water ($\mathrm{H}_2\mathrm{O}$) and carbon monoxide (CO) \citep{Barman2011,Konopacky2013,Barman2015,PetitditdelaRoche2018}. However, this data was never used to estimate the RV of the planets. 

High resolution spectroscopy of exoplanets has so far only been possible for two classes of exoplanets, hot-Jupiters and widely separated super-Jupiters. The strong and rapidly varying radial velocities of hot Jupiters can be used to isolate the planetary signal from the one of their host star, with which they are blended, as well as distinguishing them from telluric lines. The Cryogenic High-Resolution Infrared Echelle Spectrograph (CRIRES, R=100,000) at the Very Large Telescope \citep[VLT][]{Kaeufl2004} has been a prolific instrument in this field, characterizing both transiting \citep[e.g.,][]{Snellen2010} and non-transiting hot Jupiters \citep[e.g.,][]{Brogi2012}. Such measurements have enabled the detection of atomic lines \citep{Hoeijmakers2018,Salz2018}, and molecular lines (e.g. water, $\mathrm{H}_2\mathrm{O}$, and carbon monoxide, CO, \citealt{Snellen2010,Brogi2012,Rodler2012,deKok2013,Brogi2013,Birkby2013,Lockwood2014,Brogi2014,Brogi2016,Birkby2017}). HRS has also probed day side to night side winds, and the possibility of thermal inversion layers \citep{Brogi2012,Schwarz2015,Brogi2017,Nugroho2017}.
		
Current high-resolution spectrographs are not designed for high-contrast observations and are therefore limited to widely separated and bright directly imaged exoplanets.
The sensitivity of high-contrast imaging is limited by the diffracted starlight at the location of the planet, called speckles, which originate from atmospheric turbulence and optical aberrations within the instrument. The intensity of the speckles varies on large spectral scales (i.e., low spectral resolution) and can be described as a modulated stellar spectrum. The higher resolution noise is caused by photon shot-noise, detector read-noise, and imperfect modeling of the atmospheric transmission (i.e., telluric lines). The important feature of high spectral resolution instrument is that speckles can be removed with a high-pass filter while preserving the molecular signatures of the planet spectrum almost intact.

Measuring the RV of directly imaged planets is challenging due to their flux ratio with respect to the star.
The first RV measurement of an exoplanet with high-contrast imaging is $\beta$ Pictoris b. Using CRIRES observations and the cross-correlation of a carbon monoxide (CO) molecular template, \cite{Snellen2014} measured the RV ($-15.4\pm1.7\,\mathrm{km/s}$ relative to the star) and the spin of the planet ($25\pm3\,\mathrm{km/s}$). 
Observations of HR 8799 c in L-band with Keck/NIRSPEC (R = 15,000) in adaptive optics mode (NIRSPAO) provided detection of water in the atmosphere of the planet as well as a first estimate of its RV of $-8.9\pm2.5\,\mathrm{km/s}$ \citep{WangJi2018}.
Recently, H$\alpha$ was detected around two accreting exoplanets orbiting the star PDS 70, but the measured radial velocities of the emission line probe the accretion mechanism and not the orbital motion of the planet \citep{Haffert2019}.
Additionally, the RV and spin of a handful of low mass brown-dwarf companions, and larger separation or lower contrast planetary mass companions, have also been made at high spectral resolution \citep{Metchev2015,Schwarz2016,Bryan2018}.

In this work, we develop a new likelihood for the analysis of high resolution spectroscopic data. After marginalizing over the starlight modeling, we use it to measure the RV of HR 8799 b and c from Keck/OSIRIS observations.
First, in \Secref{sec:introCCF}, we discuss the current paradigm for the reduction of high resolution spectroscopic data, which is based on cross correlation.
Then, we describe the observations and data reduction in \Secref{sec:data}. The planet detection in individual exposure is described in \Secref{sec:datamodel}. The RV measurements are presented in \Secref{sec:RV}. In \Secref{sec:orbits}, the new RV data is used to better constrain the orbits of the planets.
We discuss the results and conclude in \Secref{sec:osirisdiscussion} and \Secref{sec:osirisconclusion} respectively.

A more detailed description of the calibrations and supplemental information for this chapter are provided in \Appref{app:apposirisobs}-\ref{app:transmission}. The mathematical background and derivations can be found in \Appref{app:appmaxliknD}.

\section{Preamble: the Cross Correlation Function}
\label{sec:introCCF}

RV measurement, detection of molecule, and abundance estimation from spectroscopic data are only possible with the use of atmospheric models and molecular templates. With HRS, the signal to noise ratio (S/N) of individual lines is too low to allow their independent detection and characterization. HRS therefore extensively relies on the concept of a cross correlation function (CCF), which can be seen as way to stack the signal of the individual lines together. Briefly, the data processing steps generally include the division by a transmission spectrum of the atmosphere, a high-pass filter to remove the diffracted starlight, and the cross-correlation of a template in the spectral direction to harvest the signal of the planet. The peak value of the cross correlation, expressed as a function of the RV shift of the model, is a measure of the detection strength. The shape of the cross correlation function has also been used to infer the rotational broadening of planets and brown-dwarfs \citep{Snellen2014,Schwarz2016,Bryan2018}.

Assuming Gaussian and uncorrelated noise, the cross-correlation is closely related to maximum likelihood estimation. The definition of a likelihood is important for parameter estimation (e.g. RV, spin, molecular abundance etc.) and the calculation of their uncertainty. An accurate likelihood is also paramount when combining data from different instruments with different resolutions. Indeed, the main drawback of HRS is the difficulty to constrain the local continuum used as a reference to measure the lines depths, because it is generally removed with a high-pass filter to mitigate the stellar light contamination. Low-resolution spectra or broadband photometry, on the other hand, can provide a good estimate of the continuum strength and shape. Therefore, it is desirable to jointly analyze data from different instruments, but it is seldom attempted because of the difficulty in defining the joint-likelihood.  

We define a data vector ${\bm d}$ of size $N_{\bm d}$, a centered Gaussian noise vector ${\bm n}$ with variance $\sigma^2$, and a model template ${\bm m}$, such that ${\bm d}=\epsilon {\bm m}+n$. We use the bold font convention for vectors and matrices.
If ${\bm m}_{\mathrm{RV}}$ is defined as the Doppler shifted model spectrum, the value of the cross correlation corresponding to that RV shift is given by $\mathrm{CCF(RV)} = {\bm m}_{\mathrm{RV}}^{\top}{\bm d}$. 
In the following, the subscript in ${\bm m}_{\mathrm{RV}}$ will be omitted.
Using matrix notations, the discrete cross-correlation function, expressed as a vector, can be written $\mathrm{CCF} = {\bm T}{\bm d}$, with ${\bm T}$ a rectangular Toeplitz matrix where each row is a shifted version ${\bm m}^{\top}$, e.g.:
\begin{equation}
    {\bm T} = [{\bm m}_{-1\,\mathrm{km/s}},\dots,{\bm m}_{0\,\mathrm{km/s}},\dots,{\bm m}_{+1\,\mathrm{km/s}}]^{\top}
\end{equation}
In the context of HRS, the cross correlation should not be understood as the simple mathematical operation, but as a Doppler shift for which the spectral shift is wavelength dependent.
\cite{Brogi2019} propose a new CCF to log-likelihood relationship and discuss past examples from the literature. All these mappings are statistically grounded, and only differ in their data modelling assumptions. More specifically, they differ in their choice of the free parameters, which are the amplitude of the signal, $\epsilon$, and the variance of the noise, $\sigma^2$. There are four possible cases: fitting for both, none, and one or the other parameters. First, jointly estimating $\epsilon$ and $\sigma^2$ yield the result from \citet{Zucker2003},
\begin{equation}
    \mathrm{log}(\mathcal{L}) \propto  -\frac{1}{N_{\bm d}} \mathrm{log}\left(1-\left(\frac{({\bm m}^{\top}{\bm d})}{N({\bm m}^{\top}{\bm m})({\bm d}^{\top}{\bm d})}\right)^2\right)
\end{equation}
Conversely, if both parameters are fixed, there is no free parameter to estimate and the log-likelihood is given by \citet{Lockwood2014}:
\begin{equation}
    \mathrm{log}(\mathcal{L}) \propto {\bm m}^{\top}{\bm d},
\end{equation} `
\cite{Brogi2019} only estimate the variance of the noise and assume the planet amplitude fixed, resulting in the following log-likelihood:
\begin{equation}
    \mathrm{log}(\mathcal{L}) \propto -\frac{1}{N_{\bm d}}\mathrm{log}\left( {\bm m}^{\top}{\bm m} + {\bm m}^{\top}{\bm d} - 2{\bm m}^{\top}{\bm d} \right)
\end{equation}
Finally, assuming the variance is fixed and fitting for the amplitude of the planet model yields the following likelihood \citep{Cantalloube2015,Ruffio2017}:
\begin{equation}
    \mathrm{log}(\mathcal{L}) \propto  -\frac{\left({\bm m}^{\top}{\bm d}/\sigma^2\right)^2}{{\bm m}^{\top}{\bm m}/\sigma^2},\,\,\,\,\,S/N =  \frac{{\bm m}^{\top}{\bm d}/\sigma^2}{\sqrt{{\bm m}^{\top}{\bm m}/\sigma^2}}  
    \label{eq:jbliksnrCCF}
\end{equation}

These four examples directly relate the CCF to a likelihood, but the model of the data remains simple.
We argue for a more fundamental approach to the modeling of HRS data. In statistical analysis, it is considered better practice to forward model the signal and use minimal pre-processing to avoid modifying the distribution of the noise and decrease the validity of the likelihood. 
An example of heavily processed data used for inference in HRS is the estimation of planet RV and spin from the CCF.
Assuming Gaussian noise in the original spectrum, the CCF is also expected to follow a multivariate Gaussian distribution as it is a linear function of the data, $CCF={\bm T}{\bm d}$. 
However, its values are correlated; with a characteristic correlation length given by the autocorrelation of the planet model spectrum. Note that we are here discussing the correlation of another correlation --- the CCF. Indeed, assuming homoskedastic, uncorrelated, and centered Gaussian noise for the data, the covariance matrix of the CCF values is given by $\Sigma_{\mathrm{CCF}}={\bm T}{\bm T}^{\top}/\sigma^2$. The row vectors of $\Sigma_{\mathrm{CCF}}$ are then shifted copies of the autocorrelation function of the model ${\bm m}$.
Therefore, a simple $\chi^2$ analysis of the shape of the cross correlation function will not enable robust parameter estimation.
Another more common example of data pre-processing is the interpolation of a spectrum on a regular wavelength grid, which also introduces correlated noise. Generally, any operation on the data  modifies the property of the noise, making it harder to model, and therefore limiting the accuracy of the likelihood.

\section{Observations and Data Reduction}
\label{sec:data}

\begin{figure}
  \centering
  \includegraphics[trim={0 4cm 0 3cm},clip,width=1\linewidth]{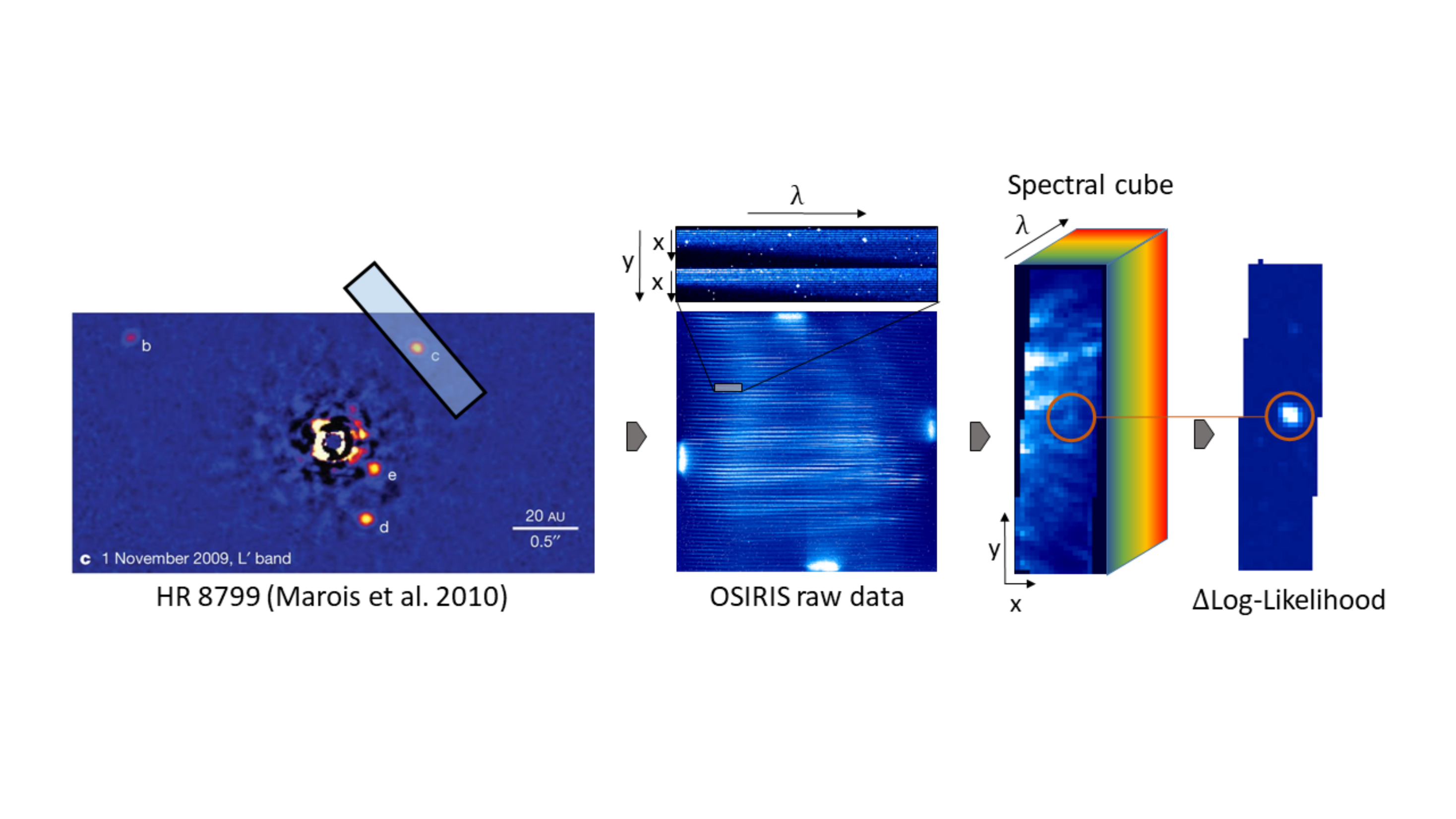}
  \caption{Schematics of the observations of the planets orbiting HR 8799 using the OSIRIS integral field spectrograph at the Keck observatory. From left to right: the HR 8799 four-planet system \citep{Marois2010} over which is laid the OSIRIS field of view (2 mas platescale mode), a raw detector image featuring the individual spaxel spectra, a sky-subtracted and mean-combined image of the reconstructed data cube from the OSIRIS data reduction pipeline, and the final detection map of a single ten minute exposure based on this work.}
  \label{fig:HR8799_OSIRIS}
\end{figure}

Over the past ten years, a total of 38 hours of observations were carried out on HR 8799 b and c in K and H-band with Keck/OSIRIS \citep{Barman2011,Konopacky2013,Barman2015,PetitditdelaRoche2018}. A description of the data and calibration acquisition strategy can be found in \citet{Barman2011}. A summary table of the relevant observations is provided in \Appref{app:apposirisobs}.
Keck/OSIRIS includes a lenslet grid and a diffraction grating providing $64\times16$ spectra at a resolution of $\sim R=4,000$. 
For all but one night, the smallest platescale of $20\,\mathrm{mas/pix}$ was used, providing a $\sim 1.3\times0.3''$ field of view (FOV), which only allows the observation of a single planet at once as shown in \autoref{fig:HR8799_OSIRIS}.
Because of mechanical uncertainties in offsetting the telescope pointing and because both the star and the planet cannot fit in the FOV, their locations is not precisely known. This requires the planet to be detected in each individual frame before the signal from different exposures can be combined. 
OSIRIS is fed by the Keck adaptive optics system, which therefore provides a diffraction limited point spread spread function. The lack of a coronagraph hurts the sensitivity to faint companions close to the star. However, the limited raw contrast is balanced by the higher spectral resolution.

OSIRIS underwent repairs and upgrades over the years changing the data quality over time \citep{Lockhart2019}. A few notable events are relevant to this work. The cooling system experienced issues during most of 2009, which significantly increased dark current \citep{Barman2011}. We have therefore excluded the 2009 epochs from this analysis.
The dispersion grating was upgraded in December 2012, which increased its efficiency by a factor $\sim2$ \citep{Mieda2014}.
In January 2016, the Hawaii-2 detector was replaced by a Hawaii-2RG \citep{Boehle2016}; greatly improving the data quality. The vast majority of the usable data has been taken prior to the upgrade.

The 1019 spectra of each spatial location (i.e., spaxel) are vertically offset from each other by only two pixels and horizontally dispersed on the detector (cf \autoref{fig:HR8799_OSIRIS}). The OSIRIS data reduction pipeline (DRP) is used to build 3D spectral cube with an iterative algorithm for the deconvolution of the overlapping spectra \citep{Krabbe2004,Lyke2017}.
The location of the spectra on the detector is stable and therefore stored in rectification matrices that are only recalibrated after hardware interventions. Rectification matrices are computed from series of white light scans where columns of lenslets are illuminated one at a time.
The wavelength solution is calculated separately using arc lamp calibrations and hard-coded in the OSIRIS DRP, which chooses the suitable file based on the date of the observation. Finally, the DRP interpolates the spectral cubes on a regular wavelength grid. 

Despite the stability of the instrument, the default wavelength solution suffers from biases.
For example, in 2017, the mean error over the field of view was estimated to be $\approx13\%$ of a pixel ($\sim5\,$km/s compared to a $\sim38\,$km/s pixel) and spatial standard deviation up to $5\%$ ($\sim2\,$km/s, \citealt{Lockhart2019}). 
In this work, we aim at measuring radial velocities at a precision and accuracy of under a kilometer per second and therefore need to correct such offsets. As prescribed in the OSIRIS pipeline user manual\footnote{\href{https://www2.keck.hawaii.edu/inst/osiris/OSIRIS\_Manual\_v2.3.pdf}{https://www2.keck.hawaii.edu/inst/osiris/OSIRIS\_Manual\_v2.3.pdf}}, we use the prominent $\mathrm{OH}^-$ radical emission lines from sky background observations as a wavelength reference (\autoref{app:wavecal}). 
To simplify its implementation, we do not derive a full third order polynomial wavelength solution such as the one used in the instrument pipeline. We instead limit the correction to a single offset per pixel and per night of observation. As a consequence, the default wavelength solution and instrument pipeline are still used to build the spectral cubes. The wavelength offset is only used in the calculation of the Doppler-shifted planet spectrum model, but it is neither accounted for in the modelling of the transmission spectrum nor the spurious starlight model. 
The calibration is described in more details in \Appref{app:wavecal}.

The transmission profile of the instrument and the atmosphere, the super-sampled point spread function (PSF), and the flux are calibrated nightly using A0 reference star observations according to \Appref{app:transmission}.

\begin{figure}[bth]
\fig{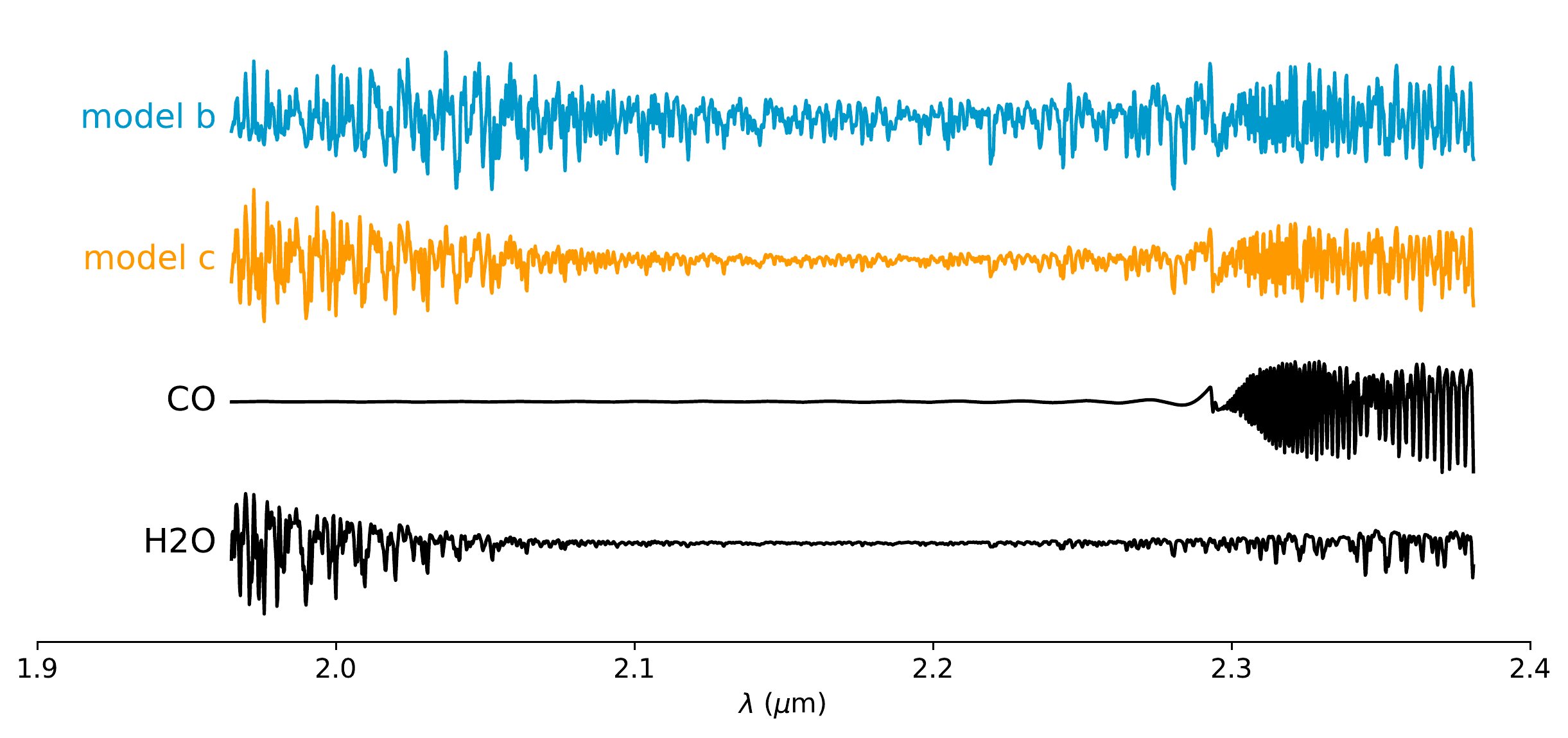}{1\textwidth}{(a) K-band}
\fig{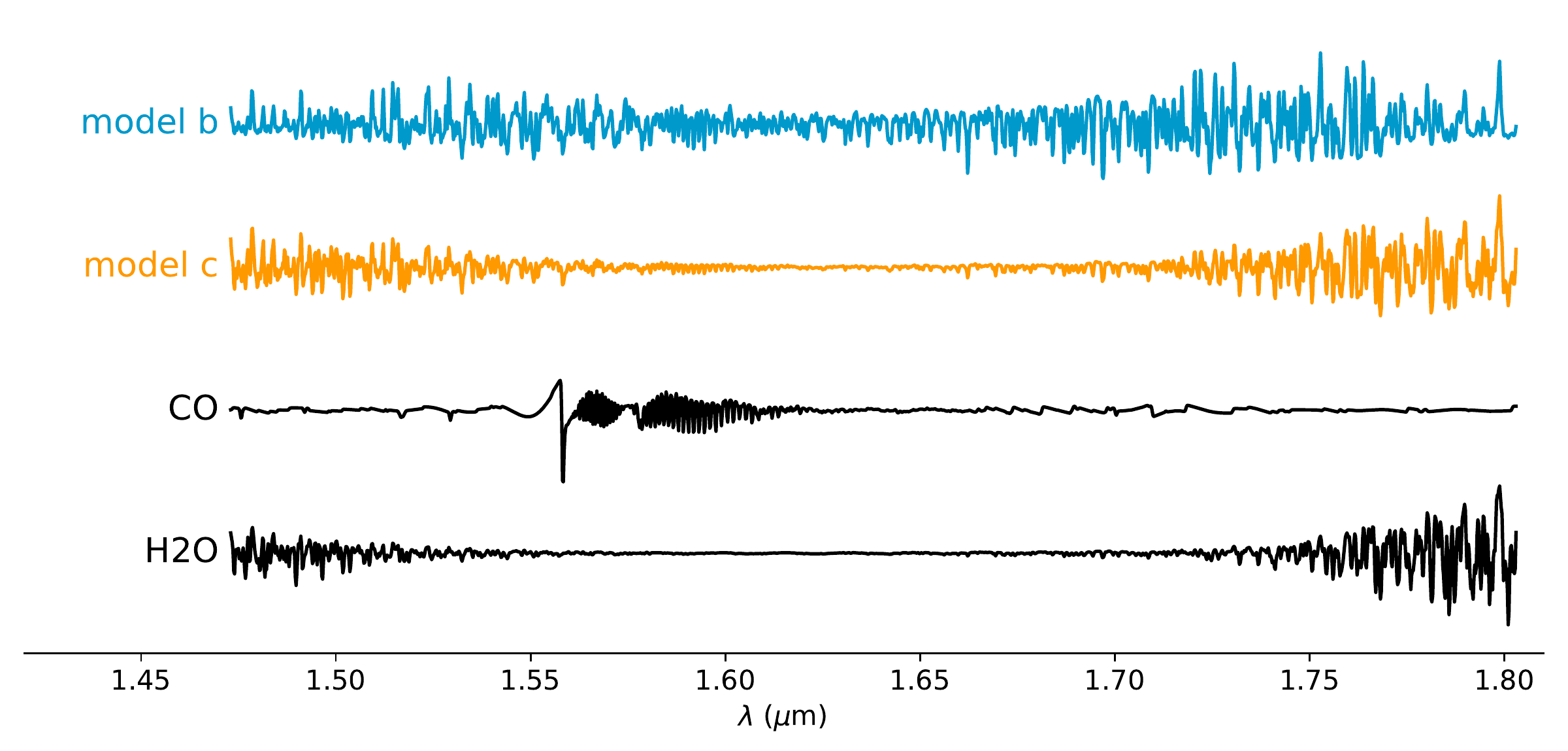}{1\textwidth}{(b) H-band}
\caption{Atmospheric and molecular templates described in \cite{Barman2011,Barman2015}. The spectra are convolved to the Keck/OSIRIS resolution ($R=4,000$) in both spectral filters: K-band (a) and H-band (b). In both figures, the first two spectra are the best fit atmospheric models resulting from \cite{Barman2015} for HR 8799 b and \cite{Konopacky2013} for HR 8799 c. These spectra are used to detect and estimate the RV of the planets. As the most prominent components of the spectra, the spectral signatures of water and carbon monoxide are shown separately, but they are not used in this work. All spectra have been normalized to a unit maximum deviation from zero and then vertically offset from each other.}
\label{fig:HR_8799_templates}
\end{figure}

A critical piece of the data analysis is the atmospheric models used for the planet detection, and RV measurement.
We use the templates described in \cite{Barman2011,Barman2015}, which are shown in \autoref{fig:HR_8799_templates} for both H and K band. The molecular templates are generated from the full atmospheric model, therefore including a realistic temperature-pressure profile, but only including a subset of opacity sources when computing the outgoing spectrum of the planet. The opacities that are accounted for are the pseudo continuum (e.g. H$_2$-H$_2$ collision induced absorption) and the specific molecule that we are trying to detect.
The spectra are convolved using a Gaussian with a wavelength dependent full width at half maximum matching the Keck/OSIRIS resolution $R=4,000$.

\section{Data model and Planet Detection}
\label{sec:datamodel}

\begin{figure}[bth]
\fig{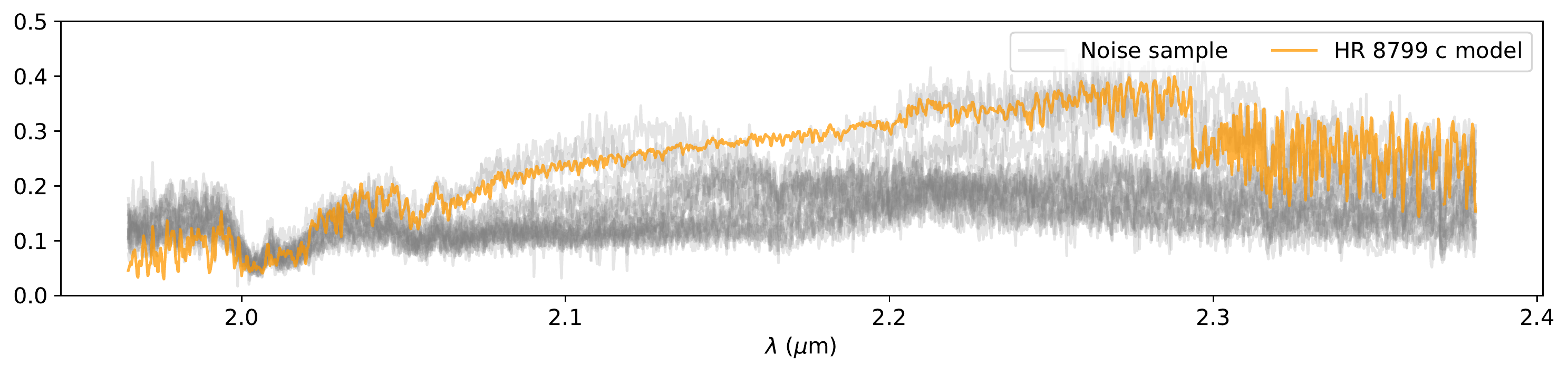}{1\textwidth}{Original}
\fig{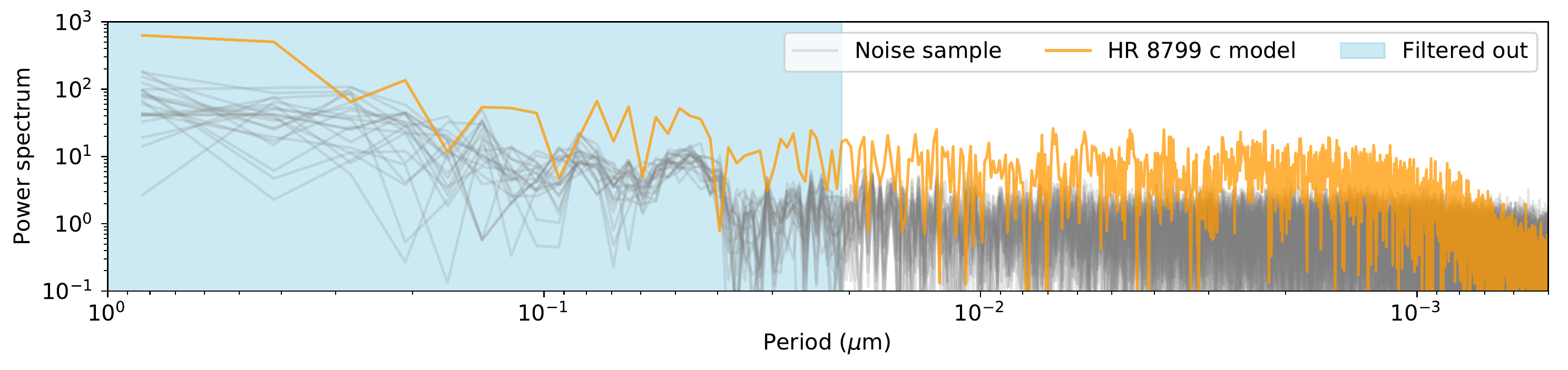}{1\textwidth}{Fourier transform}
\fig{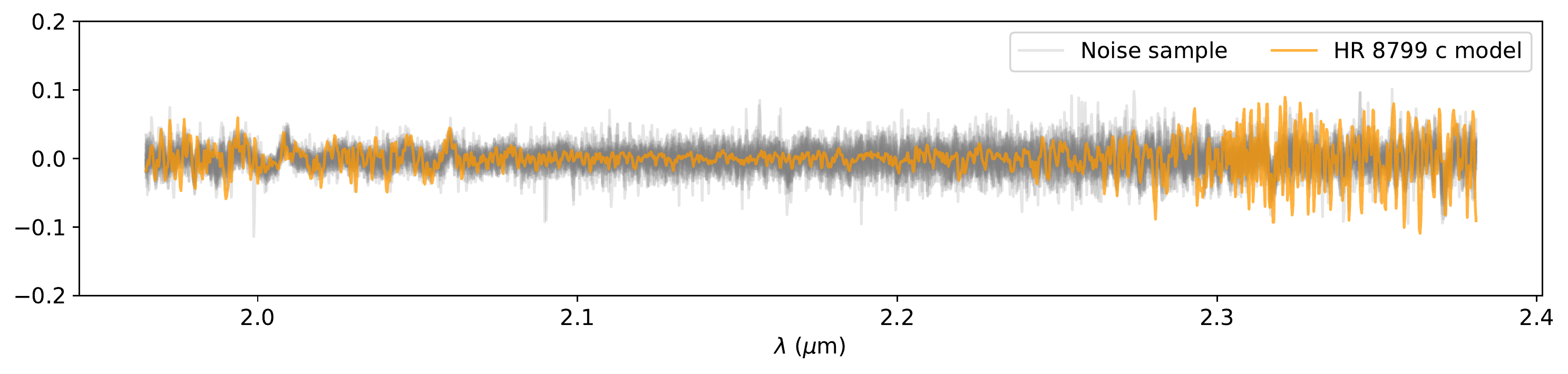}{1\textwidth}{High-pass filter}
\caption{Medium resolution spectrum ($R\approx4,000$) of a planet model compared to speckle noise from Keck/OSIRIS at K-band. We use the best fit atmospheric model of HR 8799 c from \citet{Konopacky2013}. In (a), the noise samples spectra correspond to individual spaxels in the Keck/OSIRIS field of view, which have been corrected for atmospheric and instrumental transmission using reference star observations. (b) features the Fourier transform of each spectrum, as well as the high-pass filter cutoff used in this work to subtract the speckle noise. The resulting high-pass filter spectra are illustrated in (c).}
\label{fig:HRSspeckles}
\end{figure}

The stellar halo makes the detection of planets close to the star challenging. With a $1.7''$ separation from the star, HR 8799 b can easily be seen in mean-combined raw data cube, but spotting HR 8799 c ($0.94''$) is significantly more challenging. The dithering of the image between exposures ---used to average wavelength calibration, detector noise, and sky subtraction noise--- adds another level of difficulty. Indeed, the offsets recorded in the headers are unfortunately not precise enough to allow blind stacking of the signal from different images. 
As a consequence, it is first necessary to detect and localize the planet in individual spectral cubes before any subsequent analysis. In this section, we will introduce the modeling of the data and describe the planet detection algorithm, which is based on a likelihood ratio test. The statistics background and derivations used in this work are detailed in \autoref{app:appmaxliknD}.

We chose a linear statistical model of the data ${\bm d}$, consisting of the model matrix ${\bm M}_{{\bm \psi}}$, itself a function of the parameters ${\bm \psi}$ (e.g. RV, atmospheric model etc.), linear parameters ${\bm \phi}$, and centered uncorrelated Gaussian noise vector ${\bm n}$, such that,
\begin{equation}
    {\bm d} = {\bm M}_{{\bm \psi}}{\bm \phi} + {\bm n}.
\end{equation}
Each term is described in more detailed in the following.

The data ${\bm d}$ is a vector of $N_{\bm d}=5\times5\times N_\lambda$ elements corresponding to a $5\times5$ spaxel subset of the spectral cube, including $N_\lambda$ spectral channels with $N_\lambda = 1665$ in K-band. It is a vectorized postage stamp-sized data cube centered on the assumed position of the planet. The stellar halo, or speckles, can be seen as a modulation of the stellar spectrum varying with position and wavelength. Because it is a distortion of the continuum, the effect of the speckles is minimized at higher resolution, which is why the data is generally high-pass filtered. 

Before high-pass filtering, we flag and mask bad pixels as followed. First, we retrieve the bad pixels identified by the OSIRIS data reduction pipeline and saved as fits file extensions. Then, we flag additional bad pixels by removing the pseudo continuum with a $100$-pixel median filter, and identifying outliers with a threshold equal to seven times the median absolute deviation. Immediately neighboring pixels to any bad pixels is also flagged as bad, which also include the edge of the field of view. The bad pixels are temporarily replaced by the estimated pseudo continuum.

\autoref{fig:HRSspeckles} illustrates the bad-pixel corrected data and the high-pass filtering for a single spaxel. 
We chose a linear Fourier-based high-pass filter with a cutoff corresponding to a $1/20^{\mathrm{th}}$ of the spectral band (i.e., periodicity cutoff of $21\,\mathrm{nm}$ compared the $416\,\mathrm{nm}$ width of K-band). In practice, we need to ensure continuity of the spectrum at the edge of the spectral band, which is why we first concatenate a mirrored copy of the spectrum before calculating the Fourier transform.
The cutoff can be justified from \autoref{fig:HRSspeckles}; Indeed, the noise is assumed to be uncorrelated and the Fourier transform of white noise should have a flat power spectrum. While this is approximately true at higher spectral resolution, the speckle noise introduces a lot of power at lower spectral resolution. Therefore, the cutoff is chosen to match the location where the power spectrum flattens.
The data can then be written in terms of its low and high pass filtered component so that ${\bm d}={\bm d}_{\mathrm{L}}+{\bm d}_{\mathrm{H}}$. However, to avoid cluttering the notation, we will omit the subscript and assume high-pass filtered quantities, unless specified otherwise.

The matrix ${\bm M}_{{\bm \psi}}$ includes the model of the planet as well as a model of the starlight from the host star at the location of the planet. It is therefore a function of the planet and the star spectra, the planet RV, the combined transmission of the instrument and the atmosphere, and the PSF of the instrument, which are all represented by the non-linear parameters ${\bm \psi}$. We write ${\bm M}_{{\bm \psi}}=[{\bm c}_{\mathrm{0,planet}},{\bm c}_1,\dots,{\bm c}_{25}]$, where the ${\bm c}_i$ are column vectors with the same size as the data vector ${\bm d}$ and also function of ${\bm \psi}$. The first column ${\bm c}_{0,\mathrm{planet}}$ is the high-pass filtered model of the planet, therefore equal to the vectorized PSF multiplied by the planet spectrum and the transmission, and normalized to the flux of the star. Each ${\bm c}_{i>0}$ represents the diffracted starlight of one of the 25 spaxels under the planet PSF. They are defined as,
\begin{equation}
{\bm c}_i={\bm d}_{i,L} \frac{(\mathcal{T}\mathcal{S}_{\mathrm{star}})_H}{(\mathcal{T}\mathcal{S}_{\mathrm{star}})_L}
\end{equation}
where ${\bm d}_{i,L}$ is the low-pass filtered data vector of a spaxel $i$, $\mathcal{T}$ is the mean transmission profile defined in \Secref{app:transmission}, and $\mathcal{S}_{\mathrm{star}}$ is a model spectrum of HR 8799 from the Phoenix library \cite{Husser2013}. The multiplication by ${\bm d}_{i,L}$ effectively modulates the spectral lines, such that their depths match the strength of the speckles at a given position and wavelength.

We then define the 26 linear parameters as ${\bm \phi} = [\epsilon,a_1,\dots,a_{K-1}]$, with $\epsilon$ being the planet to star flux ratio, and $a_i$ the amplitude of the starlight model ${\bm c}_i$. 

Finally, ${\bm n}$ is a centered Gaussian random vector with covariance $\Sigma=s^2\Sigma_0$, where $\Sigma_0$ is a diagonal matrix with diagonal elements equal to ${\bm d}_{L}$ and $s^2$ is a variance scaling factor.  

\begin{figure}
  \centering
  \includegraphics[width=1\linewidth]{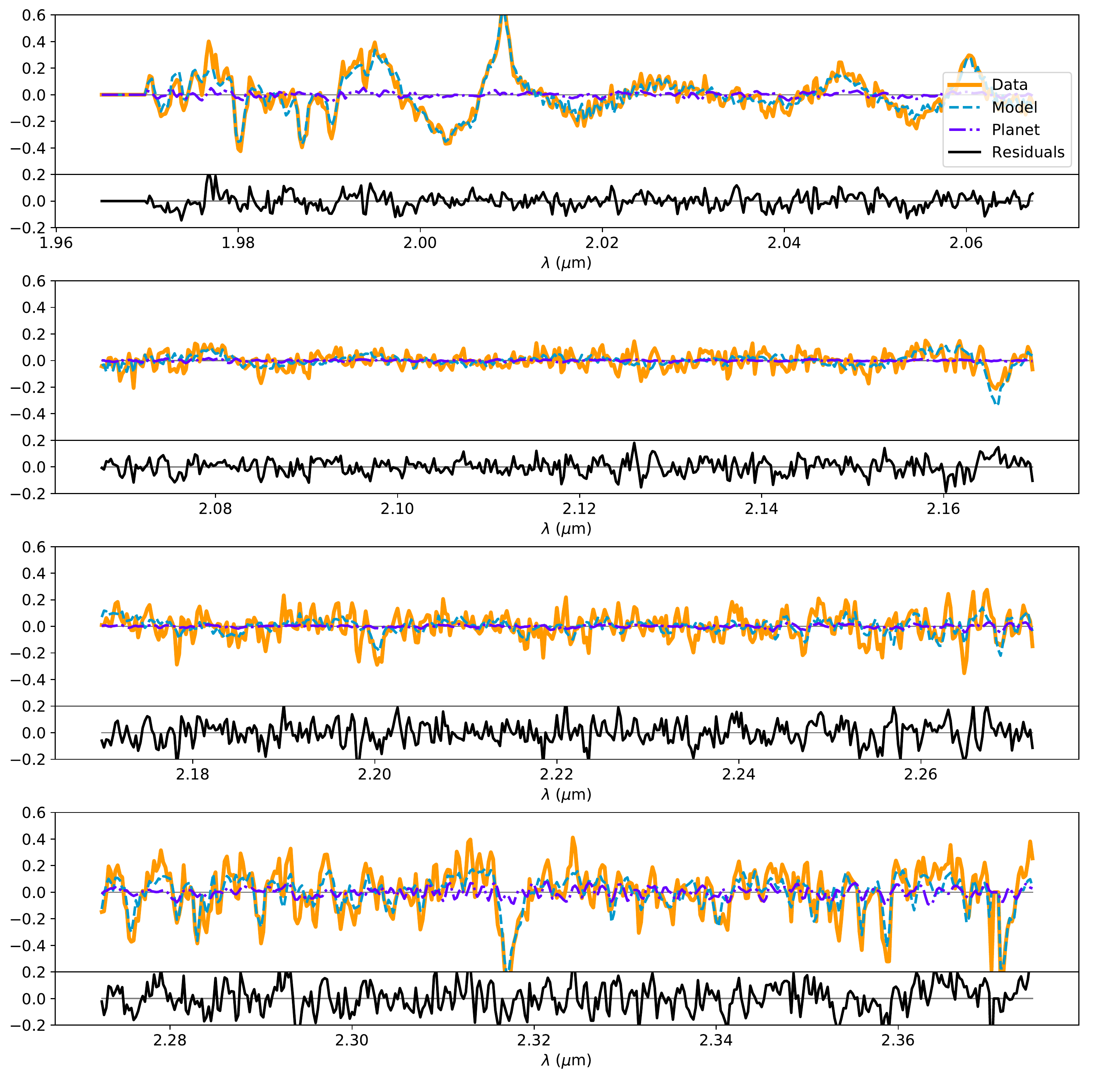}
  \caption{Residuals of a single model fit as a function of wavelength. The residuals were averaged over the $5$ pixel wide area around the location of the planet. Here, the planet model is also plotted separately from the data model, which includes both the starlight and the planet. For this figure, we used a single spectral cube of HR 8799 c.}
  \label{fig:residuals_planet}
\end{figure}

\begin{figure}
  \centering
  \includegraphics[width=1\linewidth]{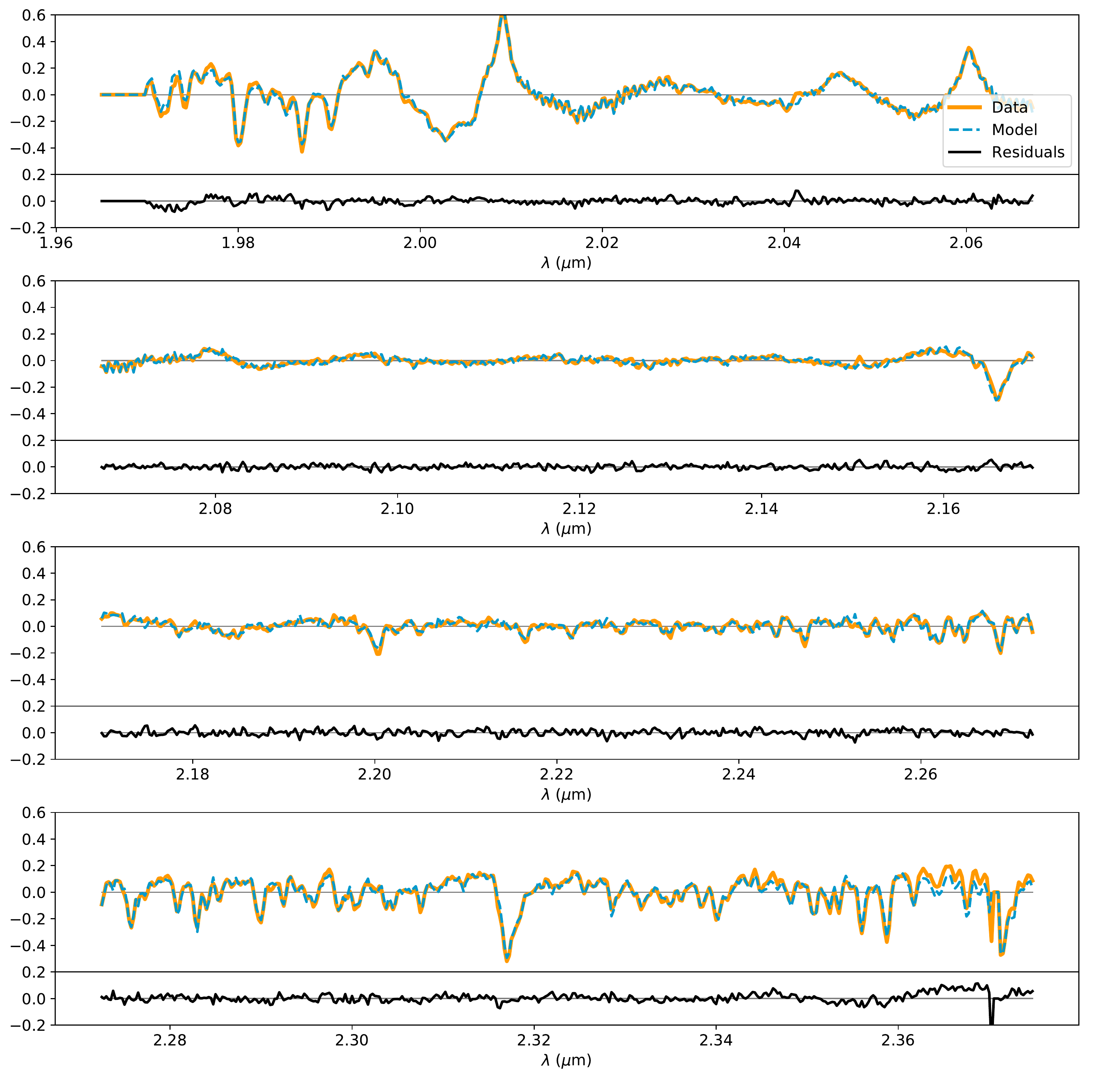}
  \caption{Average residuals over the entire field of view of a single exposure shown as a function of wavelength. The residuals were averaged over both the location of the planet and the $5$ pixel box of the data vector. For this figure, we used a single spectral cube of HR 8799 c.}
  \label{fig:residuals_avg}
\end{figure}

The likelihood is given in \Appref{app:appOSIRISmodel} and the maximum likelihood estimate of the parameters is described in \Appref{app:appmultivarmaxlik}. 
We define the maximum likelihood estimate of the linear parameters $\tilde{{\bm \phi}}$ and the maximum likelihood estimate of the covariance scaling factor $\tilde{s}$ corresponding to \autoref{eq:apppseudoinv} and \ref{eq:variancescaling} respectively. The previously identified bad pixels are masked when fitting the model.
In \autoref{fig:residuals_planet}, we show the data, model, and residuals after maximum likelihood fit, all averaged over the $5$ pixel wide area at the location of the planet. It shows that the model is a fine fit to the data. The average residuals over the entire field of view is shown in  \autoref{fig:residuals_avg}; It is calculated by fitting the planet at each location of the image.

While we assume a diagonal covariance of the noise in this work, a Cholesky decomposition could be used to to reduce the problem to a diagonal covariance matrix.
This statistical framework is flexible, and applicable to a wide variety of problems since there are no constraints on the definition of the linear model ${\bm M}$. The covariance of the estimated linear parameter and the signal to noise ratio of the planet are discussed in \Appref{app:appcov} and \Appref{app:SNRfromcov} respectively.


The planet detection consists in comparing the maximized likelihood of the data ($\mathrm{log}\mathcal{L}(\tilde{{\bm \phi}},\tilde{s}^2)$; \ref{eq:maxlikelihood}) for two different hypotheses; assuming there is a planet or not (\Appref{app:OSIRISdetection}). The planet-free model is derived from the model described previously, but omitting the planet component, which means ${\bm M}_{0{\bm \psi}}=[{\bm c}_1,\dots,{\bm c}_{25}]$ and ${\bm \phi}_0 = [a_1,\dots,a_{K-1}]$. We evaluate the logarithm of the likelihood ratio,
\begin{equation}
\mathrm{log}\mathcal{L}(\tilde{{\bm \phi}},\tilde{s}^2)-\mathrm{log}\mathcal{L}(\tilde{{\bm \phi}_0},\tilde{s}_0^2),
\end{equation}
at each location in the image and with different assumed radial velocities for the planet. The likelihood ratio is maximized at the location and for the correct RV of the planet.
The maps of log-likelihood differences at the RV of star for HR 8799 b and c are shown in \Appref{app:OSIRISdetection}. 
Because the planets have wide almost face-on orbits, their radial velocities are expected to be within a few kilometers per second of that of the star ($-12.6\pm1.4\,\mathrm{km/s}$, \citealt*{Gontcharov2006}; or $-10.9\pm0.5\,\mathrm{km/s}$, \citealt{WangJi2018}).
The barycentric correction was calculated with the Python module barycorrpy\footnote{\href{https://github.com/shbhuk/barycorrpy}{https://github.com/shbhuk/barycorrpy}} \citep{Kanodia2018ascl,Kanodia2018RNAAS}.
We then selected the best spectral cubes, with strong detections, to be used in subsequent analysis and flagged exposures with strong artifacts as bad.

\section{RV}
\label{sec:RV}

\subsection{Inference}

We propose a framework for the RV measurements of directly imaged planets and the derivation of their uncertainties. The challenge of such observations is the dominance of the stellar halo at the location of the planet. We will show how the statistical model introduced in \Secref{sec:datamodel} is well suited for such inference, and even more generally the estimation of any non-linear parameters ${\bm \psi}$. In particular, it allows the analytical marginalization of the stellar halo subtraction. We will demonstrate that the method can provide the best constraints on the RV of HR 8799 b and c using medium resolution spectroscopy ($R\approx4,000$). The H and K-bands observations prior to 2013, which represents most of the data used in this work, were already published \citep{Barman2011,Konopacky2013,Barman2015,PetitditdelaRoche2018}. However, these studies did not attempt to constrain the RV of the planets. 

In \Appref{app:nonlinesti}, we show that the posterior on ${\bm \psi}$ marginalized over the linear parameters ${\bm \Phi}$ and the covariance scaling parameter $s$ is directly related to the minimized chi-squared, $\chi^2_{{\bm \phi}=\tilde{{\bm \phi}},{\bm \psi}}$. We note that the RV of the planet is a component of ${\bm \psi}$, and we will assume that the other components of ${\bm \psi}$ are known and fixed.
Assuming an improper uniform prior for $\mathcal{P}(\mathrm{RV})$ and $\mathcal{P}({\bm \phi})$, and an improper scaling parameter prior $\mathcal{P}(s) \propto 1/s^2$, we get \autoref{eq:appRVpost},
\begin{equation}
    \mathcal{P}(\mathrm{RV} \vert {\bm d})
    \propto  \frac{1}{\sqrt{\vert{\bm \Sigma}_0\vert \times \left\vert{\bm M}_{\mathrm{RV}}^{\top}{\bm \Sigma}_0^{-1}{\bm M}_{\mathrm{RV}}\right\vert}} \left(\frac{1}{\chi^2_{{\bm \phi}=\tilde{{\bm \phi}},\mathrm{RV}}}\right)^{\frac{N_{\bm d}-N_{\bm \phi}+1}{2}}.
    \label{eq:RVpost}
\end{equation}
The $\chi^2$ is defined as,
\begin{align}
    \chi^2_{0,{\bm \phi}=\tilde{{\bm \phi}},\mathrm{RV}} &= ({\bm d}-{\bm M}\tilde{{\bm \phi}})^{\top}{\bm \Sigma}_0^{-1}({\bm d}-{\bm M}\tilde{{\bm \phi}}), \nonumber \\
    &= {\bm d}^{\top}{\bm \Sigma}_0^{-1}{\bm d} - {\bm d}^{\top}{\bm \Sigma}_0^{-1}{\bm M}({\bm M}^{\top}{\bm \Sigma_0}^{-1}{\bm M})^{-1}{\bm M}^{\top}{\bm \Sigma_0}^{-1}{\bm d}.
\end{align}
In \Secref{sec:introCCF}, we discussed how the cross correlation can be related to the log-likelihood and therefore to the $\chi^2$. Because the cross correlation can be interpreted in the context of single parameter linear model, this framework can be thought of its multidimensional generalization. Indeed, the term 
\begin{equation}
{\bm d}^{\top}{\bm \Sigma}_0^{-1}{\bm M}({\bm M}^{\top}{\bm \Sigma_0}^{-1}{\bm M})^{-1}{\bm M}^{\top}{\bm \Sigma_0}^{-1}{\bm d}, \nonumber
\end{equation}
is conceptually similar to  \autoref{eq:jbliksnrCCF},
\begin{equation}
\mathrm{log}(\mathcal{L}) \propto  \left({\bm d}^{\top}{\bm m}/\sigma^2\right) \left({\bm m}^{\top}{\bm m}/\sigma^2\right)^{-1} \left({\bm m}^{\top}{\bm d}/\sigma^2\right). \nonumber
\end{equation}

In practice, the calculation of the minimized $\chi^2$ of a linear model is straightforward and computationally efficient. As part of the planet detection step, the minimized $\chi^2$ is already calculated as a function of the RV of the planet, which can be converted into the RV posterior using \autoref{eq:RVpost}. 

\subsection{Results}

\begin{figure}
  \centering
  \includegraphics[width=1\linewidth]{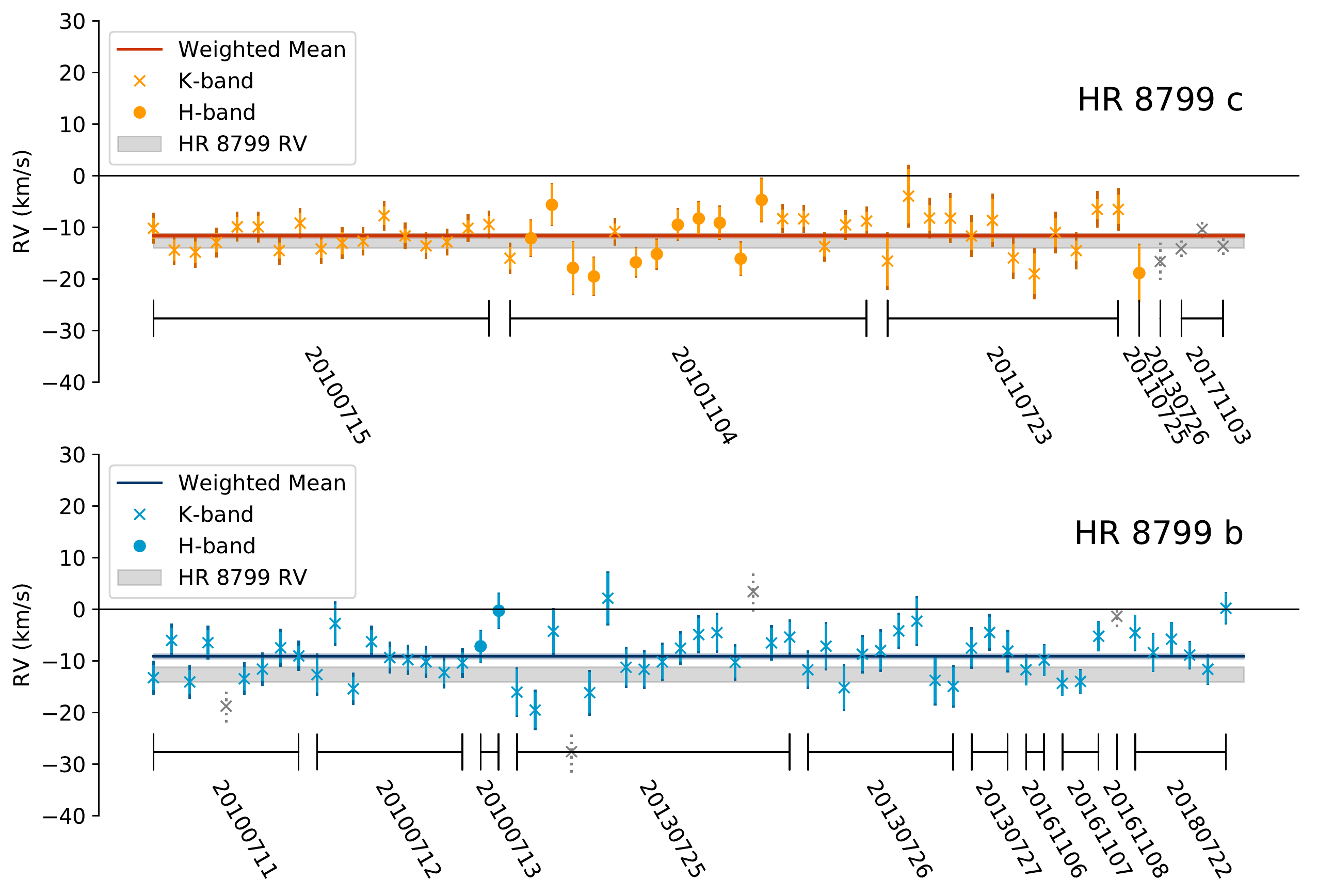}
  \caption{Radial velocity (RV) measurement of HR 8799 c (upper panel) and b (lower panel) by individual exposures. The error bars are inflated to account for the error in the wavelength solution, which is added in quadrature to the statistical error (small darker extensions of the error bars). The K and H-band observations are indicated by crosses and circle respectively. The grey regions represent the current uncertainty for the RV of the star. The greyed data points have been ignored in subsequent analysis.}
  \label{fig:RVmeasurements}
\end{figure}

The RV estimates and associated uncertainties for the selected data are shown in \autoref{fig:RVmeasurements}. Assuming coplanar and stable orbits for the HR 8799 planets, the RV of HR 8799 c, $\mathrm{RV}_\mathrm{c}$, is expected to have changed by $0.6\,\mathrm{km/s}$ since 2010 \citep{Wang2018}. We therefore only combine data spanning 2010-2011, during which the RV change is significantly below our precision. The RV of HR 8799 b, $\mathrm{RV}_\mathrm{b}$, is not expected to have changed in any measurable amount, which is why all available epochs are used. 

\begin{figure}
  \centering
  \includegraphics[trim={3cm 2cm 2cm 1.5cm},clip,width=1\linewidth]{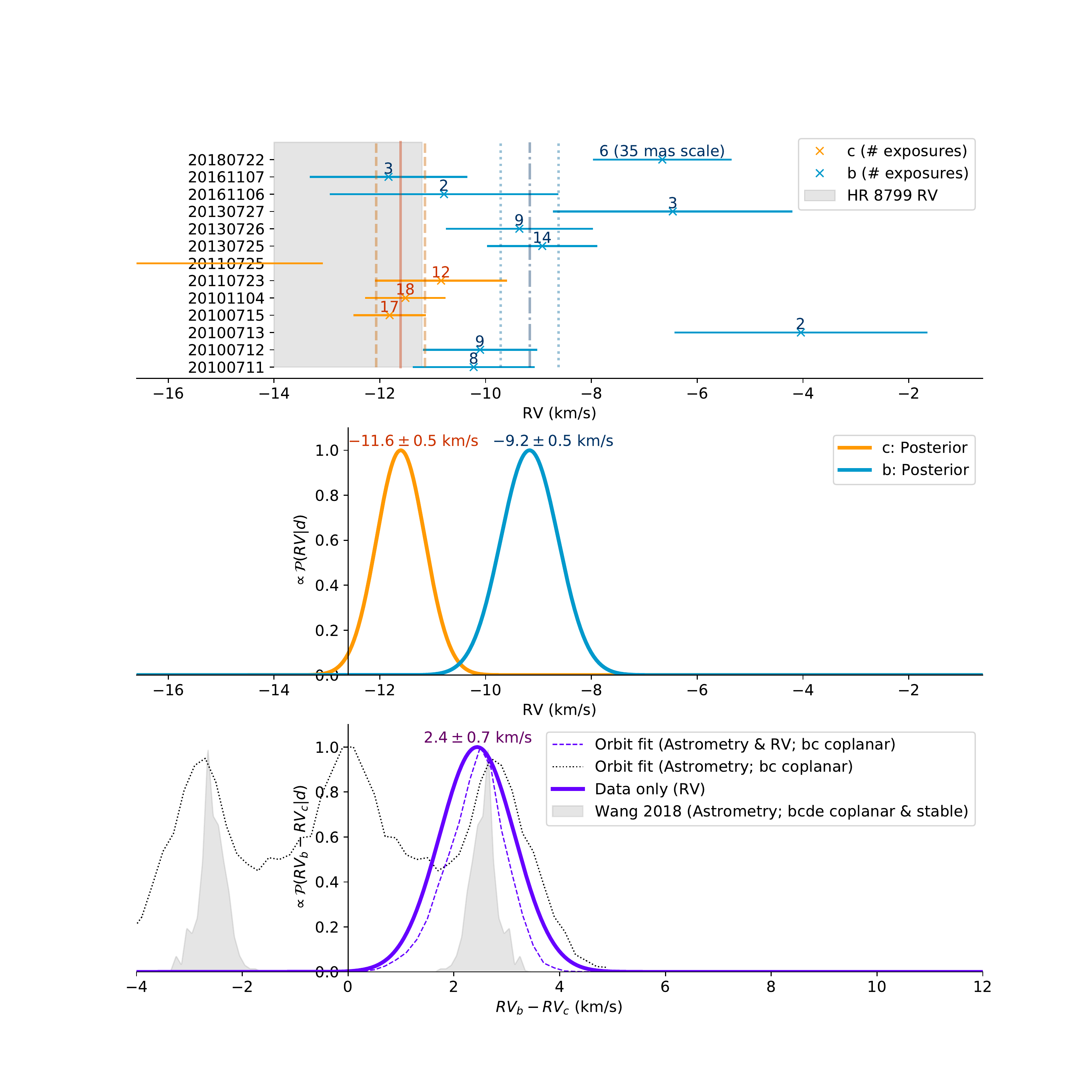}
  \caption{Combined analysis of the radial velocity (RV) measurements of HR 8799 b and c. The upper panel features the RV estimate for each individual night annotated with the number of exposures used. The middle panel shows the final posteriors when combining all data. For each planet, the tighter posterior results from multiplying individual posteriors assuming they are independent, but corrected to yield $\chi_r^2=1$. 
  The lower panel compares the posteriors on the differential RV between HR 8799 b and c resulting from a set of orbital fits. The solid purple posterior is directly derived from the middle panel. The greyed area corresponds to the posterior derived from \citep{Wang2018}, which only uses astrometric data from direct imaging and assumes that the four planets, HR 8799 b, c, d and e, are coplanar and stable. The purple dashed and black dotted posteriors are derived from the orbital fits presented in this work, with and without the RV data respectively. In both cases, the astrometric data is used and HR 8799 b and c are assumed to be coplanar, but HR 8799 e and d are not included.}
  \label{fig:RVposterior}
\end{figure}

A wavelength offset was calculated from $\mathrm{OH}^-$ emission lines in sky background observations to account for spatial and night-to-night variations of the wavelength solution. As described in \Appref{app:wavecal}, the spatial variations and the mean offset over the FOV are calculated separately. The spatial variations of the wavelength offset relative to the mean of the FOV is assumed to be constant within a year, while the mean offset is assumed to vary from night to night. An error term for the spatially varying offset is added in quadrature to the RV statistical uncertainties. The values of the error used are listed in \autoref{tab:obssummary}. Because each exposure is dithered with respect to each other, this additional error is assumed to be independent for each exposure. The inflated errors are shown in \autoref{fig:RVmeasurements} with a darker color, which suggests that the effect of the additional error is minimal compared to the statistical error. After removing $3\sigma$ outliers, the $\chi_r^2$ in \autoref{fig:RVmeasurements} for HR 8799 b and c are 1.0 and 1.4 respectively, showing fine agreement with the error bars. However, due to the larger value for HR 8799 b, we inflate the errors such as to normalize the $\chi_r^2$ to unity. The combined posteriors for $\mathrm{RV}_\mathrm{c}$ and $\mathrm{RV}_\mathrm{b}$ shown in the middle panel of \autoref{fig:RVposterior} are calculated by multiplying the individual posteriors for each exposure.

Due to the limited number of samples, it is harder to estimate the error on the nightly mean offset. The error bars in the upper panel of \autoref{fig:RVposterior} seem to already be consistent with the weighted mean of the RV, and therefore are unlikely to require an additional error term for the wavelength solution. In order to verify this hypothesis, we compare the Bayesian evidence of two models. The first model uses the uncertainties from \autoref{fig:RVposterior} to infer the mean RV of HR 8799 b. The second model includes an additional error term, which is added in quadrature to the uncertainties on the nightly estimates. When using a uniform prior for the additional error term, the Bayes factor is $3.8$ in favor of the latter model, which is not a significant difference according to \citet{Kass1995}. Additionally, the mean RV posterior marginalized over the additional error term does not significantly differ from the original posterior. We conclude that an additional calibration error is not necessary to explain the data.

As a summary, the barycentric corrected final radial velocities for HR 8799 b and c, in 2010, are estimated to be $-9.2 \pm 0.5\,\mathrm{km/s}$ and $-11.6 \pm 0.5\,\mathrm{km/s}$ respectively. The night-by-night estimates are also reported in \autoref{tab:RVepochs}. 
\begin{table}
  \centering
\begin{tabular}{ |c|ccc| } 
 \hline
  Planet & Date & RV (km/s) & \# cubes \\ 
 \hline\hline
\multirow{9}{*}{b} 
& 2010-07-11 & $-10.2 \pm 1.2$ & 8 \\
& 2010-07-12 & $-10.1 \pm 1.1$ & 9 \\
& 2010-07-13 & $-4.0 \pm 2.4$ & 2 \\
& 2013-07-25 & $-8.9 \pm 1.0$ & 14 \\
& 2013-07-26 & $-9.4 \pm 1.4$ & 9 \\
& 2013-07-27 & $-6.5 \pm 2.3$ & 3 \\
& 2016-11-06 & $-10.8 \pm 2.2$ & 2 \\
& 2016-11-07 & $-11.8 \pm 1.5$ & 3 \\
& 2018-07-22 & $-6.7 \pm 1.3$ & 6 \\
 \hline
\multirow{4}{*}{c} 
& 2010-07-15 & $-11.8 \pm 0.7$ & 17 \\
& 2010-11-04 & $-11.5 \pm 0.8$ & 18 \\
& 2011-07-23 & $-10.8 \pm 1.3$ & 12 \\
& 2011-07-25 & $-18.8 \pm 5.7$ & 1 \\
 \hline\hline
b & Combined & $-9.2 \pm 0.5$ &  \\
 \hline
c & Combined & $-11.6 \pm 0.5$ &  \\
 \hline
\end{tabular}
\caption{Barycentric corrected radial velocities of HR 8799 b and c measured with Keck/OSIRIS. For reference, the current estimate of the RV of the star is
$-12.6\pm1.4\,\mathrm{km/s}$ \citep{Gontcharov2006} or $-10.9\pm0.5\,\mathrm{km/s}$ \citep{WangJi2018}, which has not been subtracted from the values in this table due to the large uncertainty.}
\label{tab:RVepochs}
\end{table}

Due to the uncertainty on the RV of the star ($-12.6\pm1.4\,\mathrm{km/s}$; \citealt*{Gontcharov2006}; or $-10.9\pm0.5\,\mathrm{km/s}$, \citealt{WangJi2018}), it is not possible to derive tight constraints on the relative velocity of the planets to the system center of mass. However, the relative velocity of the planets with respect to each other can be used to derive the 3D orientation of the orbits.
The tightest constraints on the orbital motion of HR 8799 b and c comes from direct imaging astrometry combined with coplanarity and stability priors of the orbits \citep{Wang2018}. Because direct imaging does not distinguish between in-the-plane or out-of-the-plane of the sky motion of the planets, there exist two families of orbits with equal probability. They can be described in terms of $\mathrm{RV}_{\mathrm{b}}$ and $\mathrm{RV}_{\mathrm{c}}$ such that:
\begin{description}
\item[$1^{\mathrm{st}}$ family] $\mathrm{RV}_{\mathrm{b}}\approx+2\,\mathrm{km/s}$, $\mathrm{RV}_{\mathrm{c},2010}\approx-0.8\,\mathrm{km/s}$, and $\mathrm{RV}_{\mathrm{b}}-\mathrm{RV}_{\mathrm{c},2010}\approx+3\,\mathrm{km/s}$, 
\item[$2^{\mathrm{nd}}$ family] $\mathrm{RV}_{\mathrm{b}}\approx-2\,\mathrm{km/s}$, $\mathrm{RV}_{\mathrm{c},2010}\approx+0.8\,\mathrm{km/s}$, and $\mathrm{RV}_{\mathrm{b}}-\mathrm{RV}_{\mathrm{c},2010}\approx-3\,\mathrm{km/s}$.
\end{description}
In the bottom panel of \autoref{fig:RVposterior}, we therefore compare the posterior of $\mathrm{RV}_{\mathrm{b}}-\mathrm{RV}_{\mathrm{c},2010}$  predicted from direct imaging with the posterior derived from our RV measurements. Our result unambiguously favors the first family of orbits.

\section{Orbits fit}
\label{sec:orbits}

In the previous section, the RV predictions resulting from direct imaging assumed that the four planets orbiting HR 8799 are coplanar and stable \citep{Wang2018}.
Here, we will relax the stability constraint and study the effect of the RV on the joint orbital fit of HR 8799 b and c. The orbital parameters are estimated for two cases: with and without RV measurements.
In both cases, we assume that the two orbits are coplanar. The assumption is necessary because the radial velocity of the star is not precisely known and need to be fitted for. The relative RV of HR 8799 c with respect to the star is expected to be of the order of the measurement uncertainty, which means that its sign is not well constrained by the data. However, the sign of the relative RV of HR 8799 b is positive with high confidence. The uncertainty on the sign of HR 8799 c RV is creating two families of acceptable orbits, one of which with high mutual inclination between the two planets.
We argue that high mutual inclinations are far less likely than near coplanar orbits for stability reasons; stable orbits with mutual inclination $>8^{\circ}$ are difficult to find \citep{Wang2018}.
The two planets are forced to share the same longitude of ascending node, inclination, parallax, stellar mass, and, if applicable, stellar RV. We assume that the stellar RV and the center of mass RV are identical.
We use the RV measurements from \autoref{tab:RVepochs}, the Keck/NIRC2 astrometric measurements from \citet{Konopacky2016} (therein Table 2), and the Gemini/GPI data from \citet{Wang2018} (therein Table 2).

The posterior distribution of the parameters is calculated using the open source python package \texttt{orbitize!} \citep{blunt2019}.
Within \texttt{orbitize!}, we use the parallel tempered implementation of the affine-invariant ensemble sampler for Markov chain Monte Carlo \citep[ptemcee;][]{Vousden2016,ForemanMackey2013}. Modifications were made to Orbitize! to allow the use of planetary RVs and enforce coplanarity.

An inverse prior was used for the semi-major axis ($a\in[10^{-3},10^{7}]$). The inclination prior follows the geometric prior ($sin(i)$ with $i\in[0,\pi]$). Uniform priors were used for the eccentricity ($e\in[0,1]$), the argument of periastron ($\omega\in[0,2\pi]$), the longitude of ascending node ($\Omega\in[0,2\pi]$ if using RV, $\Omega\in[0,\pi]$ otherwise), the epoch of periastron passage ($\tau\in[0,1]$) expressed as the fraction of orbital period past November 17, 1858 at midnight (i.e., MJD=0).
Gaussian priors were used on the parallax ($24.2175\pm0.0881\,\textrm{mas}$; \citealt{Gaia2018}), the stellar mass ($1.52\pm0.15 M_\odot$; \citealt{Baines2012}) and stellar RV ($-12.6\pm1.4\,\textrm{km/s}$; \citealt{Gontcharov2006}).

The number of parameters in the fits can make the convergence of the chains difficult: 13 parameters when including the RV measurements, and 12 otherwise. Based on past work, we used 16 temperatures, 1024 walkers, and the chains were thinned by a factor 50 to remove the correlation between samples. The chains were initialized from the individual fits of the planets and run for $180,000$ steps, including a $90,000$ burn-in phase. Therefore, the results presented below include $1800$ independent samples per walker after thinning the chains ($\approx2\times10^{6}$ samples in total).

\begin{figure}
  \centering
  \includegraphics[trim={1.5cm 0cm 1.0cm 0cm},clip,width=1\linewidth]{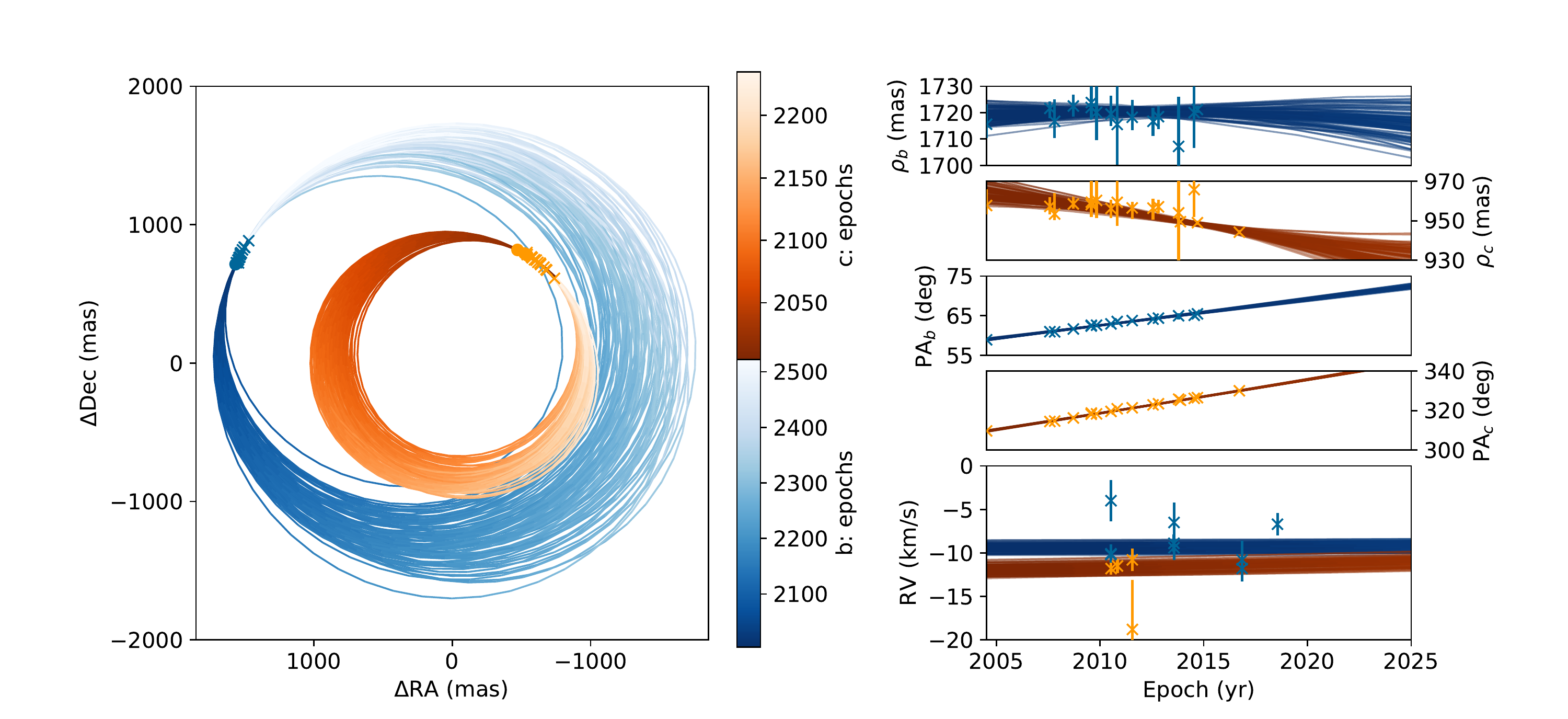}
  \caption{Orbits of HR 8799 b (blue) and c (orange) randomly sampled from their posterior. The orbital fit include the radial velocity (RV) measurements of the planets. From top to bottom on the right, the panels show the separation of b then c, the position angle of b then c, and the RV of both planets. The error bars were converted from right ascension and declination to separation and position angle using a Monte Carlo approach.}
  \label{fig:orbitsfit}
\end{figure}

\begin{figure}[bth]
\gridline{\fig{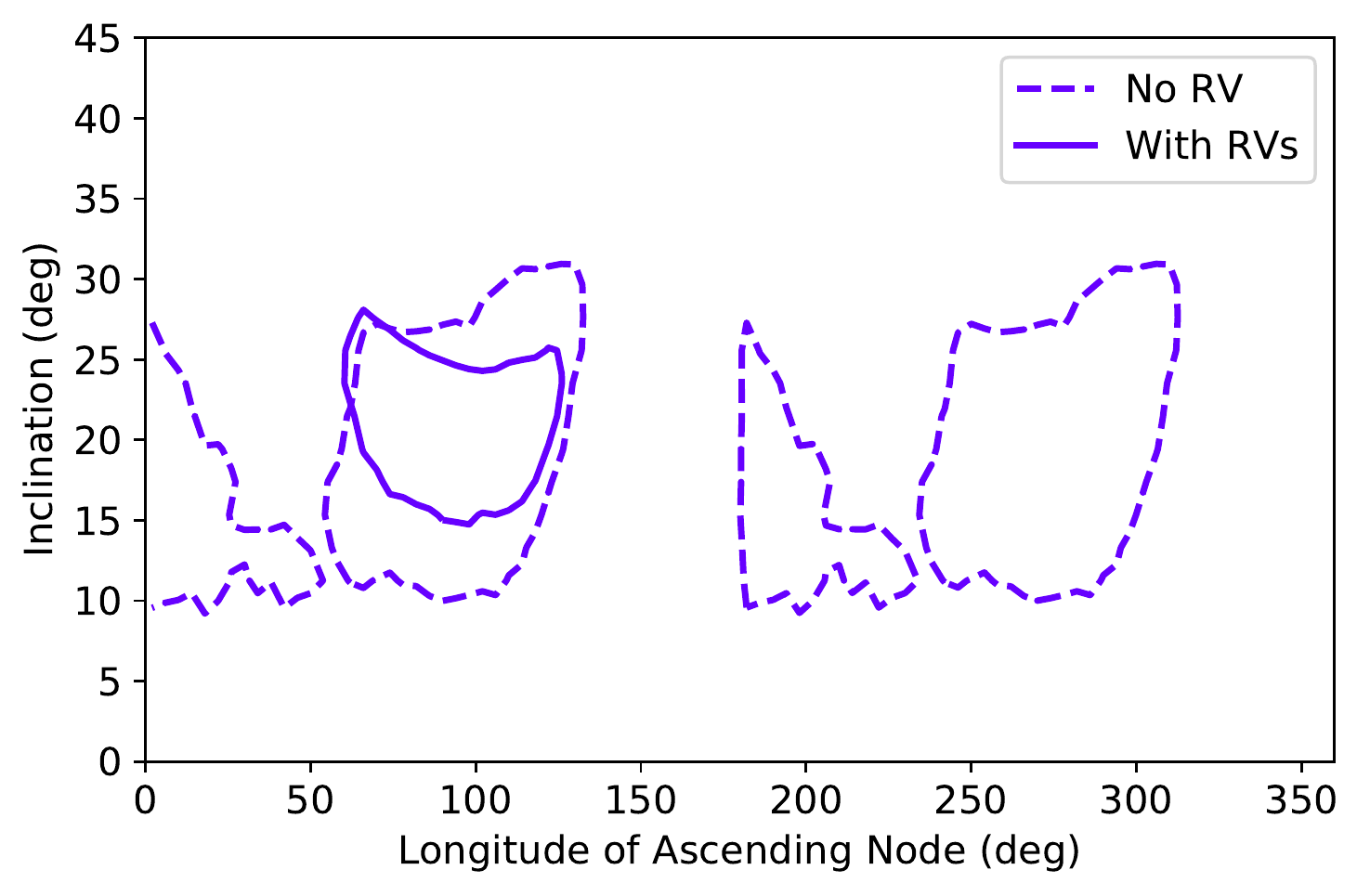}{0.49\textwidth}{(a) Orbit Orientation}
		\fig{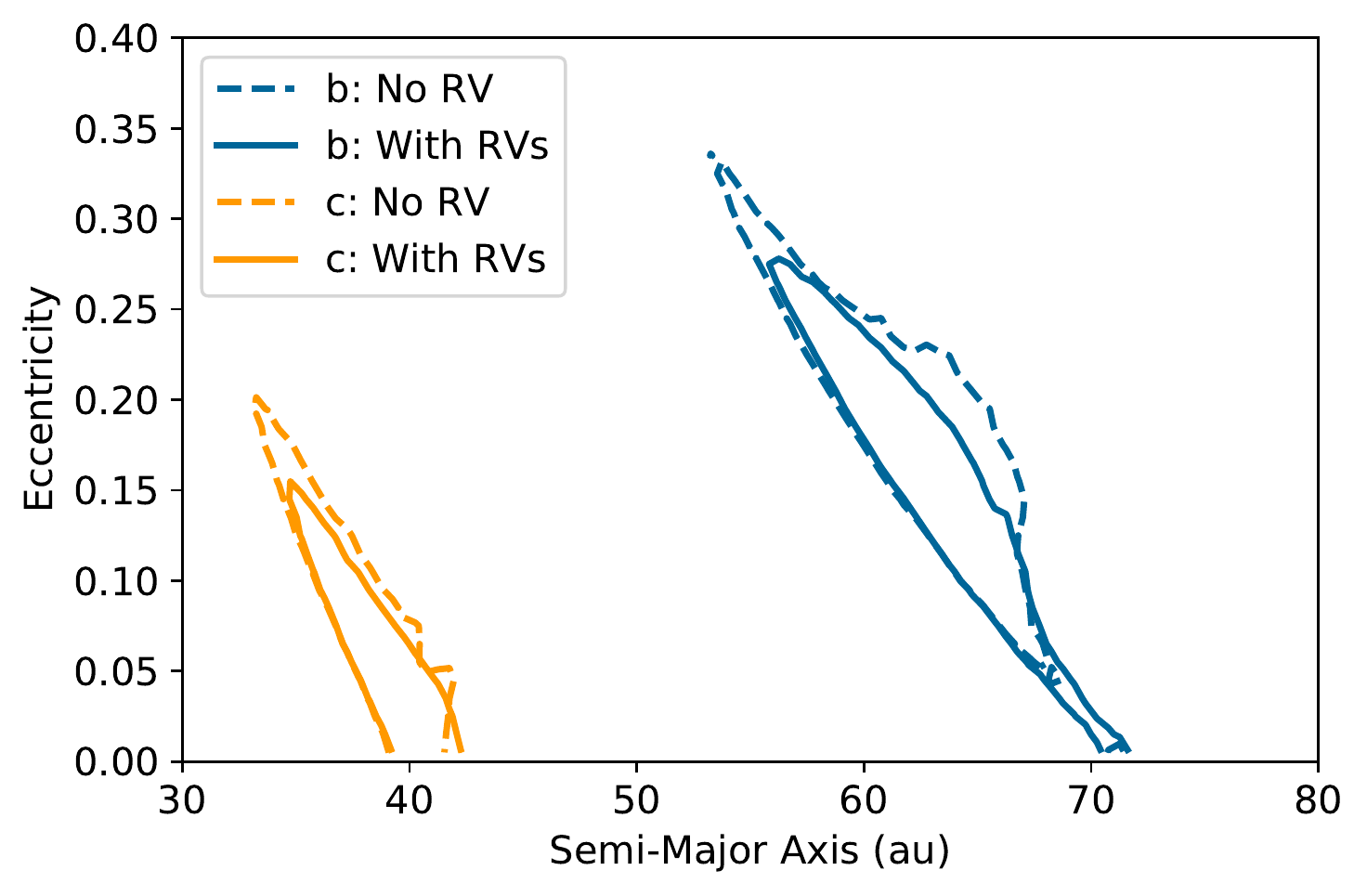}{0.49\textwidth}{(b) Orbit Shape }
          }
\caption{Contours of $68\%$ confidence interval for (a) the orientation of the orbital plane and (b) the shape of the orbit for HR 8799 b and c. The dashed contours only include the astrometric data from direct imaging, while the solid contours also include the planetary radial velocity. HR 8799 b and c are assumed to be coplanar and therefore share the same inclination and longitude of ascending node.}
\label{fig:orbitpost}
\end{figure}

For the fit including the RVs, \autoref{fig:orbitsfit} shows 50 randomly sampled orbits from the final posterior and their comparison to the data. The main advantage of the RV is to constrain the 3D orientation of the orbital plane of the planets; longitude of ascending node ($\Omega$) and inclination. However, the precision of the current measurements do not significantly constrain the shape of the orbit; semi-major axis and eccentricity. The posteriors of these four orbital parameters are shown in \autoref{fig:orbitpost} in both cases: with and without RV data. When the RV is not considered, the $\Omega$ posterior is only calculated in $[0,\pi]$, but equivalent solutions exist in $[0,2\pi]$, which is why the posterior is bi-modal. The RV unambiguously breaks down the degeneracy on $\Omega$ that exists when solely using direct imaging data. The posteriors on the differential RV between HR 8799 b and c resulting from these orbits fits is also shown in \autoref{fig:RVposterior}.

As a result, from the joint fit of HR 8799 b and c, assuming coplanar orbits and including the RV measurements, we estimate the longitude of ascending node and inclination to be $\Omega={89^{\circ}}^{\,+27}_{\,-17}$ and $i={20.8^{\circ}}^{\,+4.5}_{\,-3.7}$. As a byproduct of the fit, we can also infer the radial velocity of HR 8799 to be $\mathrm{RV}_\star =-10.5^{\,+0.5}_{\,-0.6}\,\mathrm{km/s}$, which is consistent with the value from \citet{WangJi2018} of $-10.9\pm0.5\,\mathrm{km/s}$. The derived orbital parameters are summarized in \autoref{tab:results}.

\begin{table}
  \centering
\begin{tabular}{ |cccc| } 
\hline
\multicolumn{2}{|c}{Parameters} & \multicolumn{2}{c|}{Values}\\ 
\hline\hline
$a_{b}$ && $62.3^{+4.8}_{-3.6}\,\mathrm{au}$& \\
& $a_{c}$ && $39.7^{+1.5}_{-3.1}\,\mathrm{au}$ \\
$e_{b}$ && $<0.2$& \\
&$e_{c}$  &&  $<0.09$ \\
$\omega_{b}$ && ${116.3^{\circ}}^{+42}_{-16}$& \\
&$\omega_{c}$ && ${61^{\circ}}^{+11}_{-48}$ \\
$\tau_{b}$ && $0.43^{+0.04}_{-0.06}$& \\
&$\tau_{c}$ && $0.42^{+0.10}_{-0.07}$ \\
 \hline
Parallax && $24.2^{+0.1}_{-0.1}\,\mathrm{mas}$& \\
& $M_{\mathrm{tot}}$ && $1.53^{0.11}_{-.07}M_\odot$ \\
$\mathrm{RV}_\star$ && $-10.5^{\,+0.5}_{\,-0.6}\,\mathrm{km/s}$& \\
& $\Omega$ && ${89^{\circ}}^{\,+27}_{\,-17}$ \\
$i$ && ${20.8^{\circ}}^{\,+4.5}_{\,-3.7}$ &\\
 \hline
\end{tabular}
\caption{Estimated orbital parameters from joint fit of HR 8799 b and c using planetary RVs and assuming coplanar orbits. The error bars represent the $68\%$ confidence interval. The last five parameters are shared by the two planets.}
\label{tab:results}
\end{table}

\section{Discussion}
\label{sec:osirisdiscussion}

The values of the inclination and the longitude of ascending node derived in this work remain consistent with \citet{Wang2018}: $\Omega={67.9^{\circ}}^{+5.9}_{-5.2}$ and $i=26.8^{\circ}\pm2.3$ for the stable and coplanar orbits fit using the four planets in the system. Therefore, the discussion in \citet{Wang2018} about the alignment of the planets with the debris disks remains valid; no deviations from coplanarity with mid-infrared observations with Herschel is detected \citep{Matthews2014}, but there is a slight discrepancy with longitude of ascending node of the debris disk at millimeter wavelength with the Submilimeter Array (SMA) and the Atacama Large Milimeter/submilimeter Array \citep[ALMA;][]{Wilner2018}.
A more precise measurement of the RV of the star, combined with this work, would better constrain the longitude of ascending node of the outer-planets and test the assumption of coplanarity. The tightest constraints on $\Omega$ are provided by HR 8799 d and e, which have covered longer portions of their orbital arcs. However, the two inner-planets are not included in this work.

The prospects of such measurements are exciting.
Indeed, the mass of directly imaged planets is currently estimated from atmospheric and evolutionary models, which lack absolute calibration. RV measurements of planets can contribute to their precise orbit characterization, which can enable the detection of deviations from Keplerian motion due to other planetary bodies in the system, and therefore provide independent mass measurements \citep{Lacour2019}.
Another application of RV measurements of exoplanets is the unambiguous detection of exomoons, in the very same way as RV of stars has been a powerful exoplanet detection method. To date, \cite{Teachey2018} is the only tentative detection of an exomoon around Kepler-1625b, which used transit timing and flux decrement of the star. Photo-dynamical modelling of the system shows that it is consistent with a Neptune sized moon orbiting a several Jupiter mass planet. Although the nature of the system, binary planet or exomoon is up for debate, this discovery suggests that precise doppler measurement of planets with precision under $0.1\,\mathrm{km/s}$ could already provide detections of exomoons \citep{Vanderburg2018}.

This science case will be supported by on-going and future instrument development.
For example, upgrades to NIRSPEC \citep{Martin2014SPIE}, CRIRES+ \citep{Follert2014SPIE} and the Keck Planet Imager and Characterizer (KPIC) \citep{Mawet2017SPIE} will keep opening new science opportunities. KPIC is specifically designed for high-contrast exoplanets. This new capabilities were specifically designed for high-contrast imaging and studies of planetary radial velocity or spin.
This work demonstrates that such measurements can also be made at medium spectral resolution, opening new possibilities for the James Webb space telescope. Indeed, two instruments, NIRSpec ($1-5\,\mu\mathrm{m}$, $R=1,000-2,700$) and MIRI ($5-30\,\mu\mathrm{m}$), onboard the James Webb space telescope (JWST) are mid-infrared medium resolution spectrographs very similar to Keck/OSIRIS. 

\section{Conclusion}
\label{sec:osirisconclusion}

Using data from Keck/OSIRIS, we measured the first radial velocities (RV) of HR 8799 b and c. Keck/OSIRIS is an integral field spectrograph providing $~1000$ near-infrared spectra at $R\approx4,000$ resolution over of small $\sim 1.3\times0.3''$ field of view.
Improving upon the traditional cross-correlation analysis, we developed a new forward model based statistical framework for the analysis of medium resolution spectroscopic data of directly imaged planets.
After analytically marginalizing over the starlight subtraction, we inferred the RV of HR 8799 b and c: $\textrm{RV}_b=-9.2 \pm 0.5\,\mathrm{km/s}$ and $\textrm{RV}_c=-11.6 \pm 0.5\,\mathrm{km/s}$. Despite the lower spectral resolution, these are the most precise RV measurements of directly imaged exoplanets at high-contrast. We were able to break the degeneracy in the longitude of ascending node resulting from direct imaging astrometry using the relative RV between the planets.
Assuming coplanarity, we jointly estimated the orbital parameters of the two planets and constrained the spatial orientation of the orbital plane ($\Omega={89^{\circ}}^{\,+27}_{\,-17}$ and $i={20.8^{\circ}}^{\,+4.5}_{\,-3.7}$) as well the radial velocity of HR 8799 ($\mathrm{RV}_\star =-10.5^{\,+0.5}_{\,-0.6}\,\mathrm{km/s}$).
This work demonstrates that planetary radial velocity is possible with medium resolution spectroscopy, providing exciting prospects for the James Webb space telescope (JWST).

\vspace{5mm}
\acknowledgments
{\bf Acknowledgements:} 
We would like to thank Tuan Do and Adam Mantz for their help with OSIRIS data calibration, and statistics related issues respectively.
The research was supported by grants from NSF, including AST-1411868 (J.-B.R., B.M.), 1614492 and 1614492 (T.S.B.).
Support was provided by grants from the National Aeronautics and Space Administration (NASA) NNX15AD95G (J.-B.R., R.J.D.R.). This work benefited from NASA’s Nexus for Exoplanet System Science (NExSS) research coordination network sponsored by NASA’s Science Mission Directorate.
Q.M.K., T.S.B., and K.K.W. were supported by NASA under Grant No. NNX17AB63G issued through the Astrophysics Division of the Science Mission Directorate. Any opinions, findings, and conclusions or recommendations expressed in this work are those of the author(s) and do not necessarily reflect the views of the National Aeronautics and Space Administration.
This work made use of the sky background models made available by the Gemini observatory \footnote{\url{http://www.gemini.edu/sciops/telescopes-and-sites/observing-condition-constraints/ir-background-spectra}}. 
The W. M. Keck Observatory is operated as a scientific partnership among the California Institute of Technology, the University of California, and NASA. The Keck Observatory was made possible by the generous financial support of the W. M. Keck Foundation. We also wish to recognize the very important cultural role and reverence that the summit of Mauna Kea has always had within the indigenous Hawaiian community. We are most fortunate to have the opportunity to conduct observations from this mountain.

\vspace{5mm}
\facilities{KeckI (OSIRIS)}

\software{astropy\footnote{\url{http://www.astropy.org}} \citep{Astropy2013}, Matplotlib\footnote{\url{https://matplotlib.org}} \citep{Hunter2007},
orbitize!\footnote{\url{https://github.com/sblunt/orbitize}} \citep{blunt2019},
ptemcee\footnote{\url{https://github.com/willvousden/ptemcee}} \citep{Vousden2016,ForemanMackey2013}
}

\newpage
\appendix

\section{Observations Summary Table}
\label{app:apposirisobs}

\begin{table}[h]
  \centering
\begin{tabular}{ |c|ccccccc|c| } 
 \hline
  Planet & date & band & Cubes & time & Skies & RV  & RV  & Notes \\ 
  &  &  &  & (h) &  & mean  & residual &  \\ 
  &  &  &  &  &  & offset  & error &  \\ 
  &  &  &  &  &  &  (km/s) & (km/s) &  \\ 
 \hline\hline
\multirow{19}{*}{b} 
 & 2009-07-22 & Kbb & 12 & 2.8 & 1 & 3.1 &  &   \\ 
 & 2009-07-23 & Hbb & 6 & 1.5 & 1 & 3.3 &  &   \\ 
 & 2009-07-30 & Hbb & 8 & 2.0 &  &  &  & Cooling issue  \\ 
 & 2009-09-03 & Hbb & 12 & 2.8 & 1 & -5.3 &  &   \\ 
 & 2010-07-11 & Kbb & 9 & 2.2 & 2 & 6.5 & 2.2 &   \\ 
 & 2010-07-12 & Kbb & 9 & 2.2 & 1 & 5.3 & 2.2 &   \\ 
 & 2010-07-13 & Hbb & 2 & 0.5 & 1 & 2.9 & 1.3 &   \\ 
 & 2013-07-25 & Kbb & 16 & 2.7 & 2 & 2.7 & 1.9 &   \\ 
 & 2013-07-26 & Kbb & 9 & 1.5 & 3 & 3.4 & 1.9 &   \\ 
 & 2013-07-27 & Kbb & 7 & 1.2 & 2 & 3.1 & 1.9 &   \\ 
 & 2016-11-06 & Kbb & 2 & 0.3 & 2 & 6.9 & 0.8 &   \\ 
 & 2016-11-07 & Kbb & 3 & 0.5 & 2 & 6.9 & 0.8 &   \\ 
 & 2016-11-08 & Kbb & 1 & 0.2 & 1 & 6.9 & 0.8 &   \\ 
 & 2018-07-22 & Kbb & 6 & 0.5 & 1 & -0.8 & 1.0 & $35\,\mathrm{mas}$ platescale  \\ 
\cline{2-6}
\hline
\hline
 \multirow{16}{*}{c}
 & 2010-07-15 & Kbb & 17 & 2.8 & 3 & 5.7 & 2.2 &   \\ 
 & 2010-10-28 & Hbb & 5 & 0.8 & 1 & 2.0 & 1.3 &   \\ 
 & 2010-11-04 & Hbb & 12 & 2.0 & 2 & 2.9 & 1.3 &   \\ 
 &  & Kbb & 8 & 1.3 & 2 & 5.7 & 2.2 &   \\ 
 & 2011-07-23 & Kbb & 12 & 2.0 & 3 & 3.4 & 3.0 &   \\ 
 & 2011-07-24 & Hbb & 12 & 2.0 & 3 & 0.8 & 1.5 &   \\ 
 &  & Kbb & 3 & 0.5 & 1 & 3.4 & 3.0 &   \\ 
 & 2011-07-25 & Hbb & 15 & 2.5 & 3 & 0.8 & 1.5 &   \\ 
 &  & Kbb & 2 & 0.3 & 2 & 3.4 & 3.0 &   \\ 
 & 2013-07-26 & Kbb & 1 & 0.2 &  3 & 3.4 & 1.9 &   \\ 
 & 2017-11-03 & Hbb & 14 & 2.3 & 2 & -11.4 & 1.0 &   \\ 
 &  & Kbb & 3 & 0.5 & 2 & -13.4 & 1.0 &   \\ 
\hline
\end{tabular}
\caption{Night-by-night summary of the HR 8799 observations with OSIRIS/Keck. The RV mean offset corresponds to the spatially averaged wavelength calibration offset calculated from the $\mathrm{OH}^-$ emission lines within a given night. The RV residual error is an estimated error of the spatially dependent wavelength offset relative to the spatial average (cf \Appref{app:wavecal}).}
\label{tab:obssummary}
\end{table}

\clearpage
\section{Wavelength calibration}
\label{app:wavecal}

\begin{figure}[bth]
\fig{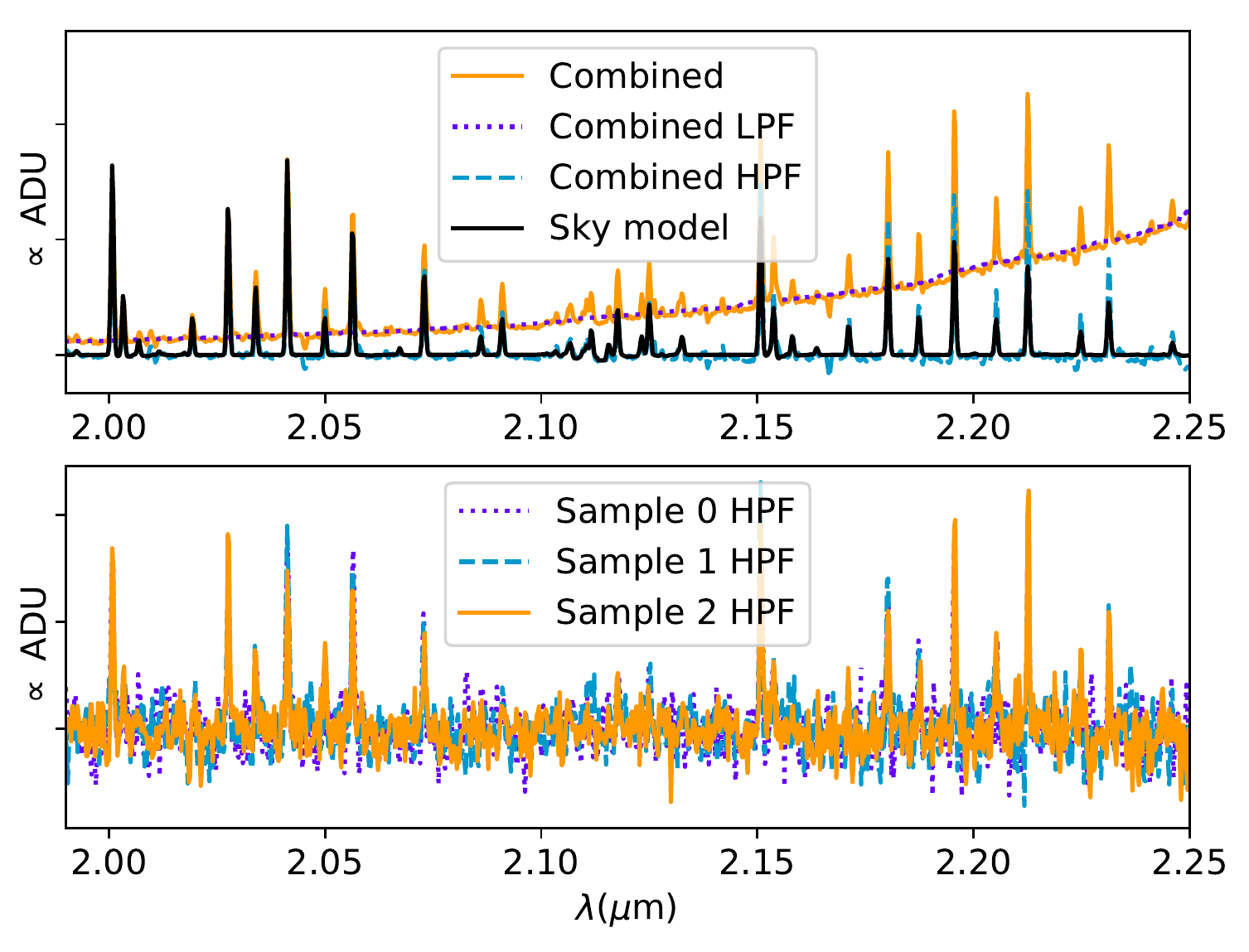}{0.59\textwidth}{(a) Sky background spectra}
\fig{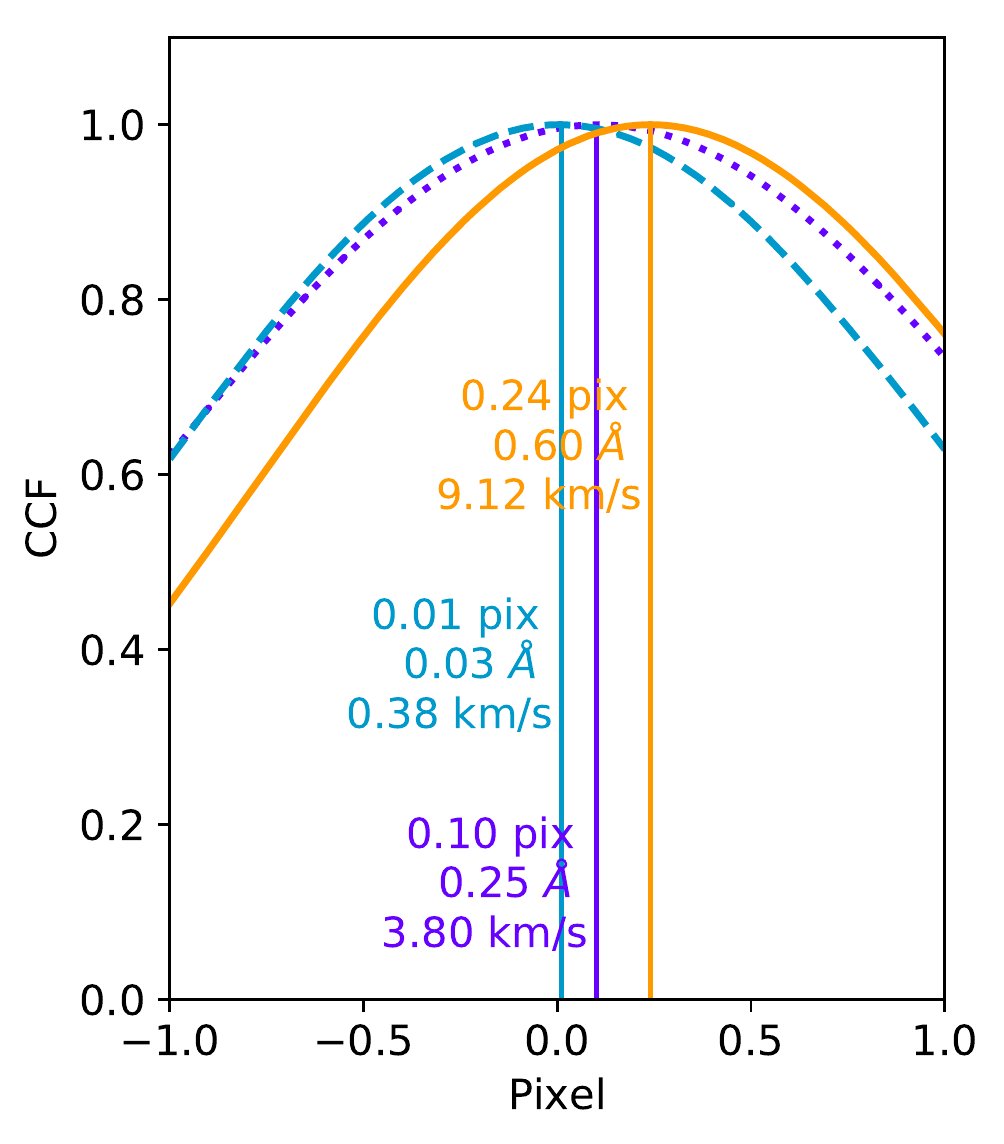}{0.39\textwidth}{(b) Cross correlation Function}
\caption{Calibration of the wavelength solution from fitting of the OH emission lines in sky background observations. The upper panel of (a) features the spatially averaged spectrum of a sky observation taken in K-band. It includes the original spectrum as well as its low-pass filtered (LPF) and high-pass filtered (HPF) components. The latter is compared to the high-pass filtered Earth atmosphere model from the Gemini observatory website. The lower panel of (a) shows three sample spectrum of individual spaxels. The right panel (b) includes the cross correlation function of the same three spectra with the model as well as the derived offsets.}
\label{fig:wavesolKbb}
\end{figure}

\begin{figure}
  \centering
  \includegraphics[width=0.59\linewidth]{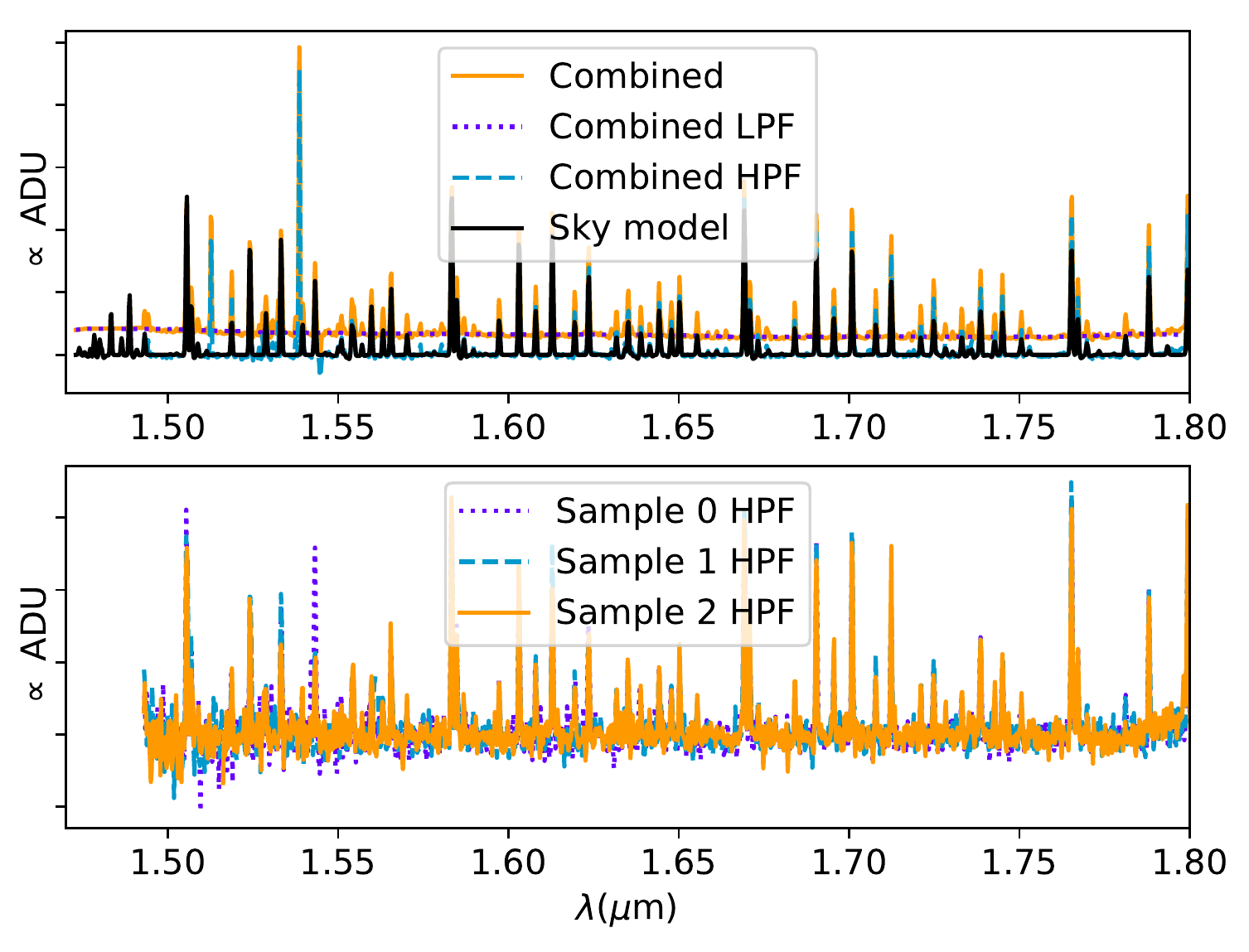}
  \caption{Similar to \autoref{fig:wavesolKbb} (a), but for H-band.}
  \label{fig:wavesolHbb}
\end{figure}

In this section, we describe the additional wavelength solution calibration used in \Secref{sec:RV}.
We use the $\mathrm{OH}^-$ radical emission lines featured in sky background observations as a wavelength reference. While there are fewer lines in K-band (\autoref{fig:wavesolKbb}) than H-band (\autoref{fig:wavesolHbb}), it is enough for calibration purposes.
To simplify its implementation, we do not derive a full third order polynomial wavelength solution such as the one used in the instrument pipeline. We instead limit the correction to a single offset, which vary as a function of pixel position and date of the observation.
The spatial variations and the mean offset over the FOV are calculated separately. The spatial variations of the wavelength offset relative to the mean of the FOV is assumed to be constant within a year, while the mean offset is assumed to vary from night to night. 

A few skies are typically acquired during each observing night and they are reduced as follow. First, each sky is dark subtracted and shaped into a three dimensional spectral cube using the OSIRIS data reduction pipeline standard recipes. We identify bad pixels in individual slices of the cube by spatially high-pass filtering them with a $5\times5$ box median filter, and using a threshold equal to seven times the median absolute deviation ($7\mathrm{MAD}$). For each spaxel, the wavelength solution offset is calculated from a maximum likelihood fit between a sky emission model spectrum and the data after they have been high-pass filtered in the spectral direction (\autoref{fig:wavesolKbb}-\ref{fig:wavesolHbb}).
The model of the sky were downloaded from the Gemini observatory website\footnote{\href{http://www.gemini.edu/sciops/telescopes-and-sites/observing-condition-constraints/ir-background-spectra}{http://www.gemini.edu/sciops/telescopes-and-sites/observing-condition-constraints/ir-background-spectra}} \citep{Lord1992}, convolved with a Gaussian matching the resolution of the instrument (Full Width at half maximum corresponding to $R=4000$), shifted by a wavelength offset, evaluated on the OSIRIS wavelength grid, and high-pass filtered before being fitted to the data.
Due to the sharpness of the lines, we use a 40-pixel wide median high-pass filter instead of the Fourier-based high-pass filtered used elsewhere in this work.

\begin{figure}
  \centering
  \includegraphics[width=1\linewidth]{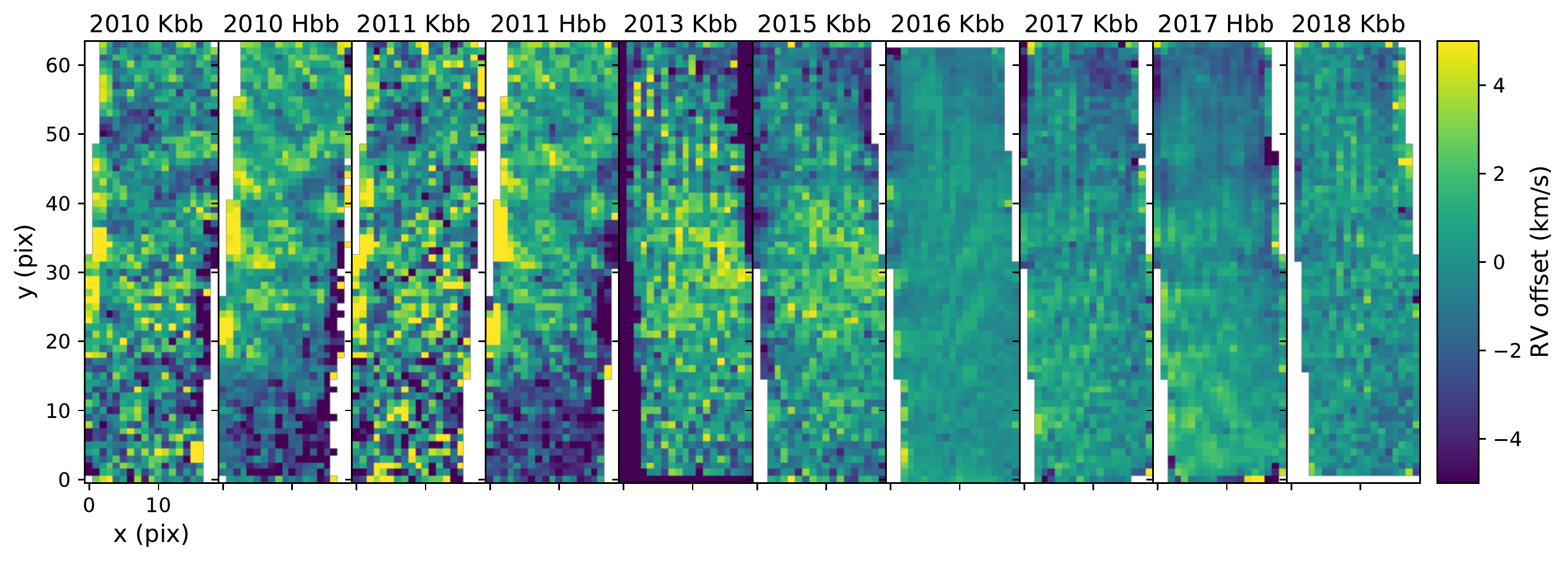}
  \caption{Spatial variation of the wavelength solution offset calculated for each year and spectral band.}
  \label{fig:wavecal}
\end{figure}

We then use a matched filter to find the optimal wavelength offset for each spaxel in the FOV. The matched filter is here identical to a cross correlation since we use a solid wavelength offset instead of a Doppler shift, and also because we assume a covariance matrix equal to the identity.
As a result of this process, a spatial map of wavelength offsets is obtained for each sky observation in our dataset. For each year, we then derive master map of the spatial variations by averaging the mean-subtracted maps. This spatial calibration maps for each year are shown in \autoref{fig:wavecal}. The error on the spatial offset calibration is calculated for each sky from the residual map after subtracting the mean-corrected master map. We report the smallest standard deviation for each year in \autoref{tab:obssummary}. This error is used to inflate the RV measurement error for each individual exposure.
The mean offset is computed by taking the average of all the wavelength offsets maps taken during a single night. The resulting mean offset for each night is given in \autoref{tab:obssummary}, which show deviations from the instrument calibration up to $13.4\,$km/s in 2017.

The master calibration for each night is calculated by adding the corresponding mean offset, calculated for each night, to the map of spatial variation, calculated for each year.

\clearpage
\section{Transmission and PSF calibration}
\label{app:transmission}

In this section, we describe the calibration of the transmission profile of the atmosphere, the flux calibration, and the derivation of the point spread function (PSF) of the instrument, which are all based on reference star observations.

\begin{figure}[bth]
\fig{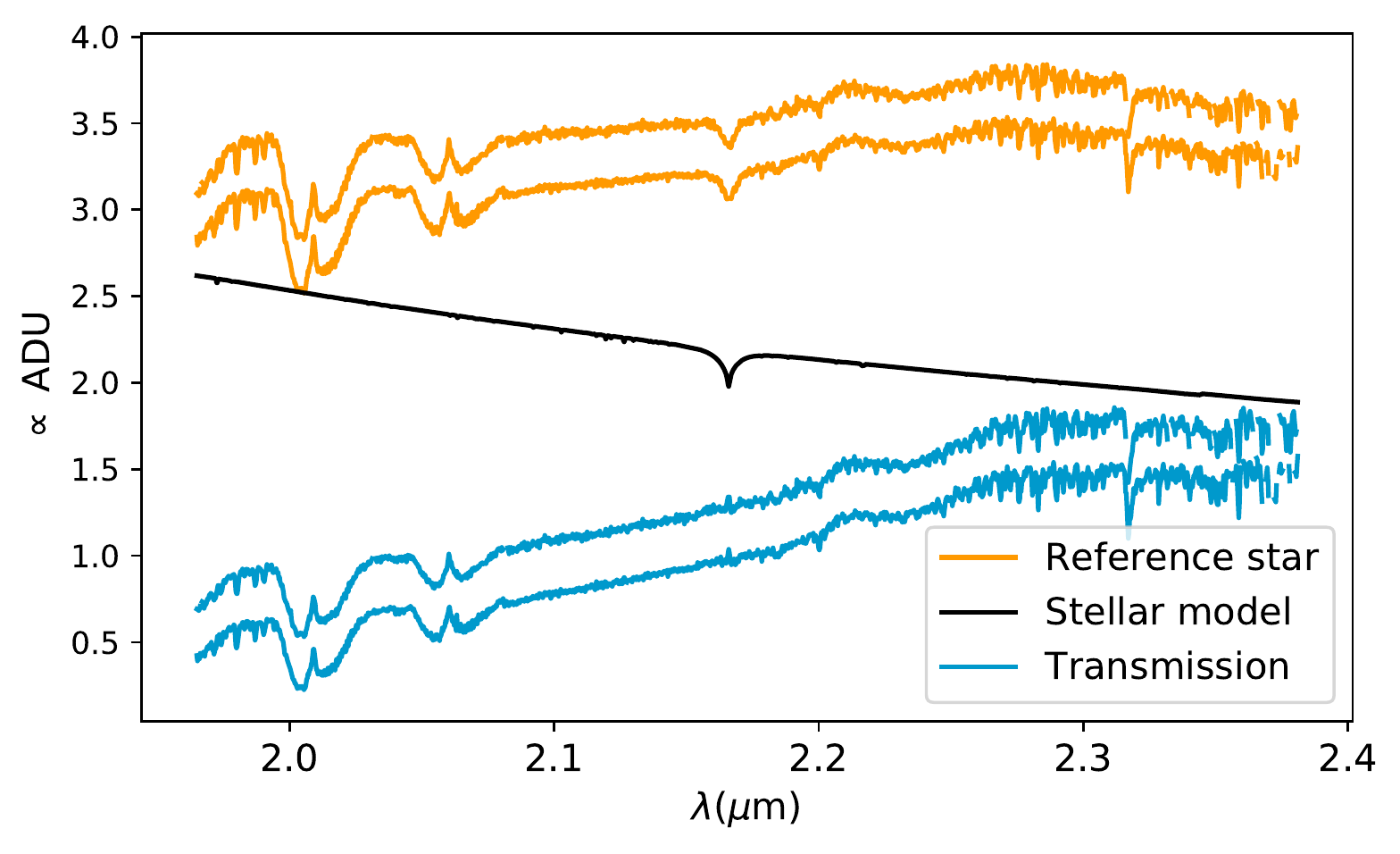}{0.59\textwidth}{(a) Transmission calculation}
\fig{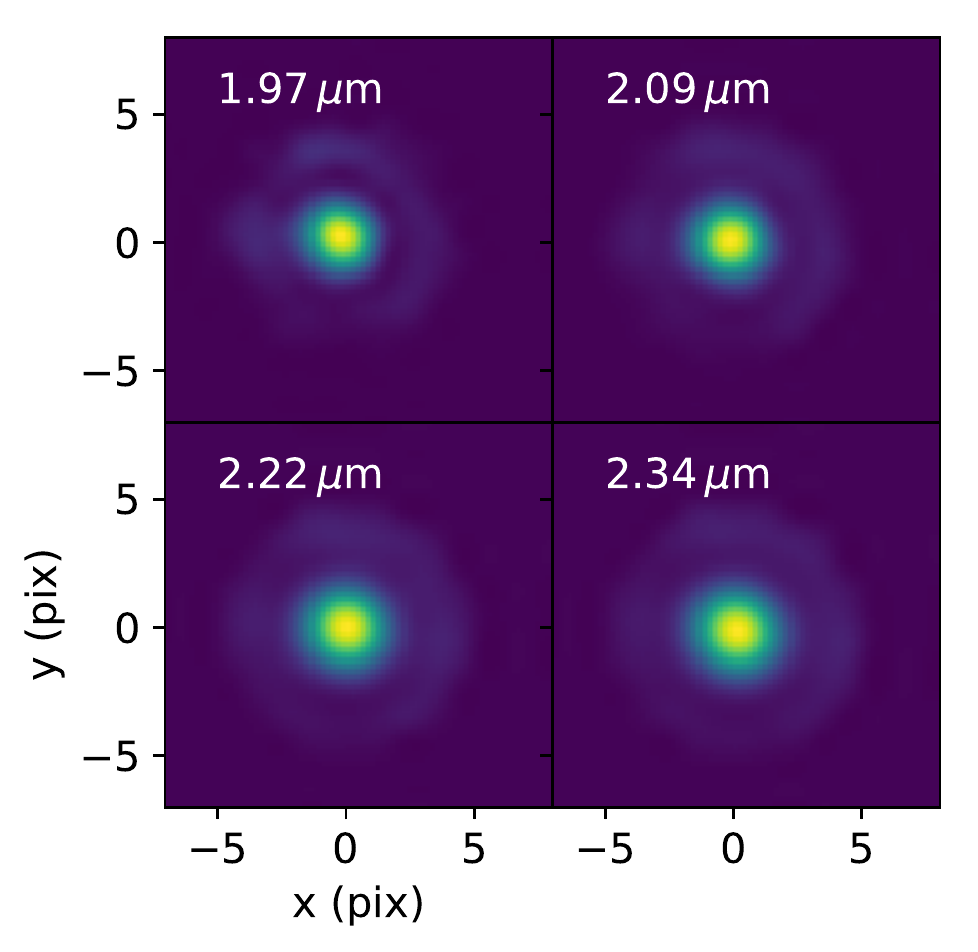}{0.39\textwidth}{(b) Super-sampled PSF}
\caption{Calibration of the transmission profile and super-sampled point spread function (PSF) from reference star observations. (a) The left panel shows the spectra extracted from two different reference star observations in K-band, the Phoenix stellar model, and the resulting tranmission spectrum. (b) The right panel shows the super-sampled PSF calculated for one night at four different wavelengths in K-band.}
\label{fig:refstar}
\end{figure}

To avoid saturating the detector, the reference stars were sometimes acquired in open loop with no adaptive optics (AO) correction. Only the AO corrected observations were used to derive the PSF and the flux calibration, while all were used to derive the transmission profile. Instead of a sky subtraction, we subtract each reference star observation in pair thanks to the dithering pattern. Bad pixels are identified in the spectral direction using a 100 pixel box median high-pass filter and a $7\mathrm{MAD}$ threshold. The neighboring pixels of such identified bad pixels are also marked as bad and subsequently masked.
The centroid of the stellar PSF is calculated in each slice from a two-dimensional Gaussian fit. Despite the atmospheric dispersion correction from the instrument, the centroid of the star varies by half a pixel in H-band, which needs to be accounted for when modelling the planet signal.
The reference center is chosen to be the median centroid of the star across the spectral band.
At each wavelength, we build a spline interpolated model of the super sampled PSF by combining the reference star observations for each spectral band on a nightly basis. An example of super-sampled PSF is given in \autoref{fig:refstar}.
For the AO corrected exposures, the star spectrum is then simply calculated by summing the flux in a five-pixel radius aperture. When the AO was turned off, we mask out the lower tenth percentile of pixels, replace the bad pixels by the median value in the given slice and integrate the resulting cube in the spatial direction to derive the spectrum.

The transmission spectrum of the atmosphere and of the instrument is calculated on a nightly basis by dividing the reference star spectrum with a Phoenix stellar model \citep[\autoref{fig:refstar};][]{Husser2013}\footnote{\url{http://phoenix.astro.physik.uni-goettingen.de}}. The stellar models are broaden using the Python module PyAstronomy\footnote{\url{https://www.hs.uni-hamburg.de/DE/Ins/Per/Czesla/PyA/PyA/index.html}} to account for the spin of the star, convolved to OSIRIS resolution and evaluated onto the data wavelength grid.

\clearpage
\section{Statistics of Multivariate Linear Model}
\label{app:appmaxliknD}

\subsection{Data Model}
\label{app:appOSIRISmodel}
We define ${\bm d}$ as the data vector of size $N$ representing pixel values and ${\bm n}$ as a Gaussia nrandom vector with zero mean and covariance matrix ${\bm \Sigma}=s^2{\bm \Sigma}_0$; $s^2$ being a scaling parameter. We define a linear model of the data with a matrix ${\bm M}$ of size $N\times N_{\bm \phi}$. The $N_{\bm \phi}$ linear parameters are denoted by the vector ${\bm \phi}$. Additionally, we can assume that the model is a function of non-linear parameters ${\bm \psi}$, such that the data model can be written,
\begin{equation}
    {\bm d} = {\bm M}_{{\bm \psi}}{\bm \phi} + {\bm n}.
\end{equation}
More specifically, the linear parameters ${\bm \phi}$ represents the amplitude of the planet and the amplitude of the stellar light at any given pixel, while ${\bm \psi}$ includes parameters defining the atmospheric model of the planet, its radial velocity, and its spin.

The corresponding likelihood is given by
\begin{align}
    \mathcal{L}({\bm \psi},{\bm \phi},s^2) &= \mathcal{P}({\bm d} \vert {\bm \psi},{\bm \phi},s) ,\nonumber \\
    &= \frac{1}{\sqrt{(2\pi)^{N}\vert{\bm \Sigma}\vert}} \exp\left\lbrace-\frac{1}{2}({\bm d}-{\bm M}_{{\bm \psi}}{\bm \phi})^{\top}{\bm \Sigma}^{-1}({\bm d}-{\bm M}_{{\bm \psi}}{\bm \phi})\right\rbrace ,\nonumber \\
    &= \frac{1}{\sqrt{(2\pi)^{N}\vert{\bm \Sigma}_0\vert s^{2N}}} \exp\left\lbrace-\frac{1}{2s^2}({\bm d}-{\bm M}_{{\bm \psi}}{\bm \phi})^{\top}{\bm \Sigma}_0^{-1}({\bm d}-{\bm M}_{{\bm \psi}}{\bm \phi})\right\rbrace.
\end{align}

\subsection{Maximum likelihood estimate}
\label{app:appmultivarmaxlik}

The most likely value of ${\bm \phi}$, noted $\tilde{{\bm \phi}}$, for given values of $s^2$ and ${\bm \psi}$, is calculated by minimizing the negative log-likelihood. The subscript in ${\bm M}_{{\bm \psi}}$ will be dropped in this section, because ${\bm \psi}$ is assumed to be fixed.
\begin{align}
    \tilde{{\bm \phi}} &= \textrm{Argmax}_{\bm \phi} \, \mathcal{L}({\bm \psi},{\bm \phi},s^2), \nonumber \\
    &=\textrm{Argmin}_{\bm \phi}  \, -\mathrm{log} \mathcal{L}({\bm \psi},{\bm \phi},s^2) .
\end{align}

The negative log-likelihood is given by:
\begin{align}
    -\mathrm{log}\mathcal{L} &=  \frac{N}{2}\mathrm{log}(2\pi) + \frac{1}{2}\mathrm{log}(\vert{\bm \Sigma}_0\vert) +\frac{N}{2}\mathrm{log}s^2 + \frac{1}{2s^2}({\bm d}-{\bm M}{\bm \phi})^{\top}{\bm \Sigma}_0^{-1}({\bm d}-{\bm M}{\bm \phi}), \nonumber \\
    &= \frac{N}{2}\mathrm{log}(2\pi) + \frac{1}{2}\mathrm{log}(\vert{\bm \Sigma}_0\vert) +\frac{N}{2}\mathrm{log}s^2 + \frac{1}{2s^2}\chi_0^2,
\end{align}
where we defined $\chi_0^2=s^2\chi^2$ as,
\begin{align}
    \chi_0^2 &= ({\bm d}-{\bm M}{\bm \phi})^{\top}{\bm \Sigma}_0^{-1}({\bm d}-{\bm M}{\bm \phi}), \nonumber \\
    &= {\bm d}^{\top} {\bm \Sigma}^{-1} {\bm d} +{\bm \phi}^{\top} {\bm M}^{\top} {\bm \Sigma}^{-1} {\bm M}{\bm \phi} - 2 {\bm \phi}^{\top} {\bm M}^{\top} {\bm \Sigma}^{-1} {\bm d}. 
\end{align}

The solution to a linear $\chi^2$ minimization problem is the pseudo-inverse.
\begin{equation}
\label{eq:apppseudoinv}
\tilde{{\bm \phi}} = ({\bm M}^{\top}{\bm \Sigma_0}^{-1}{\bm M})^{-1}{\bm M}^{\top}{\bm \Sigma_0}^{-1}{\bm d}
\end{equation}
Indeed, let's solve for $\nabla_{\bm \phi}\chi_0^2 = 0$.
\begin{align}
    \nabla_{\bm \phi}\chi_0^2 = 0 &\Leftrightarrow 2{\bm M}^{\top} {\bm \Sigma}_0^{-1} {\bm M}\tilde{{\bm \phi}} - 2{\bm M}^{\top} {\bm \Sigma}_0^{-1} {\bm d} = 0, \nonumber \\
    &\Leftrightarrow{\bm M}^{\top} {\bm \Sigma}_0^{-1} {\bm M}\tilde{{\bm \phi}} = {\bm M}^{\top} {\bm \Sigma}_0^{-1} {\bm d}.
\end{align}
Multiplying by the inverse of ${\bm M}^{\top}{\bm \Sigma}_0^{-1}{\bm M}$ returns \autoref{eq:apppseudoinv}. In practice, that inversion is never performed because it is algorithmicly much faster and more stable to solve the system of linear equations directly rather than inverting the matrix.

We can also jointly optimize for the covariance scaling factor $s^2$, which corresponds to the variance of the noise if ${\bm \Sigma}_0$ is the identity matrix. The optimization remains convex, so there is still a unique solution. The optimal $s^2$ can be derived from minimizing the profile negative log-likelihood $\mathcal{L}(\tilde{{\bm \phi}},s^2)$,
\begin{align}
    \left.\frac{\diff  (-\mathrm{log}\mathcal{L})}{\diff {\bm s^2}}\right\vert_{{\bm \phi}=\tilde{{\bm \phi}}} = 0 &\Leftrightarrow  \frac{N}{2\tilde{s}^2} - \frac{1}{2\tilde{s}^4} \chi^2_{0,{\bm \phi}=\tilde{{\bm \phi}}} = 0, \nonumber \\
    &\Leftrightarrow \tilde{s}^2 = \frac{1}{N} \chi^2_{0,{\bm \phi}=\tilde{{\bm \phi}}}.
    \label{eq:variancescaling}
\end{align}

\subsection{Covariance of the linear parameters}
\label{app:appcov}

In this section, we will calculate the covariance of the estimated linear parameters $\tilde{{\bm \phi}}$, which is given in \autoref{eq:apppseudoinv},
\begin{align}
\tilde{{\bm \phi}} &= ({\bm M}^{\top}{\bm \Sigma_0}^{-1}{\bm M})^{-1}{\bm M}^{\top}{\bm \Sigma_0}^{-1}{\bm d}, \nonumber \\
&= ({\bm M}^{\top}{\bm \Sigma_0}^{-1}{\bm M})^{-1}{\bm M}^{\top}{\bm \Sigma_0}^{-1}({\bm M}{\bm \phi}_{\textrm{true}} + {\bm n}).
\end{align}
In the second line, we replaced the data vector by its signal and noise components.

The linear transformation of a Gaussian random vector remains Gaussian, so the posterior of $\tilde{{\bm \phi}}$ must be Gaussian too. Indeed, if ${\bm X}\sim\mathcal{N}({\bm \mu_x},{\bm \Sigma_x})$, then ${\bm Y}={\bm A}{\bm X}$ also follows a normal distribution with vector mean,
\begin{equation}
    {\bm \mu_y} = {\bm A}{\bm \mu_x}
\end{equation}
and covariance matrix,
\begin{equation}
    {\bm \Sigma_y} = {\bm A}{\bm \Sigma_x}{\bm A}^{\top}.
\end{equation}
The proof is as follow,
\begin{align}
    {\bm \Sigma_y} &= \textrm{E}\left( \left({\bm y}-{\bm \mu_y}\right) \left({\bm y}-{\bm \mu_y}\right)^{\top} \right), \nonumber \\
    &= \textrm{E}\left( {\bm y}{\bm y}^{\top}-{\bm \mu_y}{\bm y}^{\top} -{\bm y}{\bm \mu_y}^{\top}+{\bm \mu_y}{\bm \mu_y}^{\top} \right), \nonumber \\
    &= \textrm{E}\left( {\bm y}{\bm y}^{\top}\right)-{\bm \mu_y}{\bm \mu_y}^{\top} , \nonumber \\
    &= \textrm{E}\left( \left({\bm A}{\bm x}\right) \left({\bm A}{\bm x}\right)^{\top} \right) -\left({\bm A}{\bm \mu_x}\right)\left({\bm A}{\bm \mu_x}\right)^{\top}, \nonumber \\
    &= {\bm A}\textrm{E}\left( {\bm x}^{\top}{\bm x}\right){\bm A}^{\top}-{\bm A}{\bm \mu_x}^{\top}{\bm \mu_x}{\bm A}^{\top}, \nonumber \\
    &={\bm A}{\bm \Sigma_x}{\bm A}^{\top}.
\end{align}

As a result, the covariance matrix of $\tilde{{\bm \phi}}$ is given by,
\begin{align}
    \textrm{cov}(\tilde{{\bm \phi}}) &= \left[ ({\bm M}^{\top}{\bm \Sigma_0}^{-1}{\bm M})^{-1}{\bm M}^{\top}{\bm \Sigma_0}^{-1}\right]s^2{\bm \Sigma_0}\left[ ({\bm M}^{\top}{\bm \Sigma_0}^{-1}{\bm M})^{-1}{\bm M}^{\top}{\bm \Sigma_0}^{-1}\right]^{\top}, \nonumber \\
    &=s^2 ({\bm M}^{\top}{\bm \Sigma_0}^{-1}{\bm M})^{-1}{\bm M}^{\top}{\bm \Sigma_0}^{-1} {\bm \Sigma_0}{\bm \Sigma_0}^{-1}{\bm M}({\bm M}^{\top}{\bm \Sigma_0}^{-1}{\bm M})^{-1}, \nonumber \\
    &=s^2 ({\bm M}^{\top}{\bm \Sigma_0}^{-1}{\bm M})^{-1}{\bm M}^{\top}{\bm \Sigma_0}^{-1}{\bm M}({\bm M}^{\top}{\bm \Sigma_0}^{-1}{\bm M})^{-1}, \nonumber \\
    &=s^2 ({\bm M}^{\top}{\bm \Sigma_0}^{-1}{\bm M})^{-1}.
    \label{eq:appcov}
\end{align}

Another way to show this result is to consider the likelihood, and write ${\bm \phi} = \tilde{{\bm \phi}}+\Delta{\bm \phi}$
\begin{align}
    \mathcal{L} &= \frac{1}{\sqrt{(2\pi)^{N}\vert{\bm \Sigma}_0\vert s^{2N}}} && \exp\left\lbrace-\frac{1}{2s^2}({\bm d}-{\bm M}{\bm \phi})^{\top}{\bm \Sigma}_0^{-1}({\bm d}-{\bm M}{\bm \phi})\right\rbrace, \nonumber \\
    &= \frac{1}{\sqrt{(2\pi)^{N}\vert{\bm \Sigma}_0\vert s^{2N}}}  && \exp\left\lbrace-\frac{1}{2s^2}({\bm d}-{\bm M}\tilde{{\bm \phi}}-{\bm M}\Delta{\bm \phi})^{\top}{\bm \Sigma}_0^{-1}({\bm d}-{\bm M}\tilde{{\bm \phi}}-{\bm M}\Delta{\bm \phi})\right\rbrace, \nonumber \\
    &= \frac{1}{\sqrt{(2\pi)^{N}\vert{\bm \Sigma}_0\vert s^{2N}}} && \exp\left\lbrace-\frac{1}{2s^2}({\bm d}-{\bm M}\tilde{{\bm \phi}})^{\top}{\bm \Sigma}_0^{-1}({\bm d}-{\bm M}\tilde{{\bm \phi}})\right\rbrace, \nonumber \\
    &&&*\exp\left\lbrace-\frac{1}{2s^2}({\bm M}\Delta{\bm \phi})^{\top}{\bm \Sigma}_0^{-1}({\bm M}\Delta{\bm \phi})\right\rbrace, \nonumber \\
    &&&*\exp\left\lbrace+\frac{1}{s^2}({\bm d}-{\bm M}\tilde{{\bm \phi}})^{\top}{\bm \Sigma}_0^{-1}({\bm M}\Delta{\bm \phi})\right\rbrace, \nonumber \\
    &= \frac{1}{\sqrt{(2\pi)^{N}\vert{\bm \Sigma}_0\vert s^{2N}}} && \exp\left\lbrace-\frac{1}{2s^2}\chi^2_{0,{\bm \phi}=\tilde{{\bm \phi}}}\right\rbrace\exp\left\lbrace-\frac{1}{2s^2}\Delta{\bm \phi}^{\top}({\bm M}^{\top}{\bm \Sigma}_0^{-1}{\bm M})\Delta{\bm \phi}\right\rbrace.
    \label{eq:applikdphi}
\end{align}
In the third equality, we used the fact that the argument of last exponential term vanishes because,
\begin{align}
    \nabla_{\bm \phi}\chi_0^2 = 0 &\Leftrightarrow 2{\bm M}^{\top} {\bm \Sigma}_0^{-1} {\bm M}\tilde{{\bm \phi}} - 2{\bm M}^{\top} {\bm \Sigma}_0^{-1} {\bm d} = 0, \nonumber \\
    &\Leftrightarrow {\bm M}^{\top} {\bm \Sigma}_0^{-1} ({\bm d}-{\bm M}\tilde{{\bm \phi}}) = 0, \nonumber \\
    &\Leftrightarrow ({\bm d}-{\bm M}\tilde{{\bm \phi}})^{\top}{\bm \Sigma}_0^{-1}{\bm M}=0.
\end{align}
Therefore,
\begin{equation}
    \mathcal{L} \propto \exp\left\lbrace-\frac{1}{2s^2}\Delta{\bm \phi}^{\top}({\bm M}^{\top}{\bm \Sigma}_0^{-1}{\bm M})\Delta{\bm \phi}\right\rbrace,
\end{equation}
which shows that the posterior of ${\bm \phi}$ with uniform priors is Gaussian with covariance,
\begin{equation}
\textrm{cov}(\tilde{{\bm \phi}}) = s^2({\bm M}^{\top}{\bm \Sigma}_0^{-1}{\bm M})^{-1}.
\end{equation}

\subsection{Signal to Noise Ratio}
\label{app:SNRfromcov}

If the planet signal corresponds to the first column of the model ${\bm \phi}$, which is assumed to be case, then the standard deviation is the square root of the first diagonal element of $\textrm{cov}(\tilde{{\bm \phi}})$. So the theoretical signal-to-noise ratio of the planet can be written as,
\begin{equation}
\mathcal{S}  = \frac{ \tilde{{\bm \phi}}[0] }{ \sqrt{ \textrm{cov}(\tilde{{\bm \phi}})[0,0]}}.
\label{eq:appgeneralizedSNR}
\end{equation}
This expression includes a marginalization over the all the other linear parameters of ${\bm \phi}$.

\subsection{Planet detection}
\label{app:OSIRISdetection}

We first define two hypotheses, $\mathcal{H}_0$ and $\mathcal{H}_1$, which correspond to the null hypothesis and the planet hypothesis respectively,
\begin{equation}
    {\bm d} = {\bm M}_{{\bm \psi}0}{\bm \phi}_{0} + {\bm n}\,\,\, \mathrm{and} \,\,\, {\bm d} = {\bm M}_{{\bm \psi}1}{\bm \phi}_{1} + {\bm n}.
\end{equation}
Assuming the null hypothesis, the only signal in the data is the diffracted starlight from the star (i.e., speckles). The planet case is simply modeled from the null hypothesis by adding one extra parameter to model the planet signal. As a result, ${\bm M}_{{\bm \psi}1}$ has one extra column, which includes the spectrum and the PSF of the planet, compared to ${\bm M}_{{\bm \psi}0}$.

When jointly fitting for ${\bm \phi}$ and $s^2$, the minimized negative log-likelihood is given by,
\begin{align}
    -\mathrm{log}\mathcal{L}(\tilde{{\bm \phi}},\tilde{s}^2) &= \frac{N}{2}\mathrm{log}(2\pi)+\frac{1}{2}\mathrm{log}(\vert{\bm \Sigma}_0\vert) +  \frac{N}{2}\mathrm{log}\tilde{s^2}+\frac{1}{2\tilde{s^2}}\chi^2_{0,{\bm \phi}=\tilde{{\bm \phi}}}  ,\nonumber \\ 
    &=\frac{N}{2}\mathrm{log}(2\pi)+\frac{1}{2}\mathrm{log}(\vert{\bm \Sigma}_0\vert) +  \frac{N}{2}\mathrm{log}\tilde{s^2}+\frac{1}{2\tilde{s^2}}\chi^2_{0,{\bm \phi}=\tilde{{\bm \phi}}}  ,\nonumber \\
    -2\mathrm{log}\mathcal{L}(\tilde{{\bm \phi}},\tilde{s}^2) &=N\mathrm{log}(2\pi)+ \mathrm{log}\vert{\bm \Sigma}_0\vert + N\mathrm{log}\chi^2_{0,{\bm \phi}=\tilde{{\bm \phi}}} - N\mathrm{log}N + N.
    \label{eq:maxlikelihood}
\end{align}

There are different methods to estimate the likelihood of one model compared to another. A few common examples are: likelihood ratio, the Akaike information criterion (AIC), or the Bayesian information criterion (BIC). AIC and BIC include a penalty for extra model parameters. This can be useful if one wants to modeled the planet signal as linear combination of molecular templates for example. However the penalty for extra parameters is a constant, which means that for our simple planet detection scheme, a likelihood ratio is satisfactory.

\begin{figure}
  \centering
  \includegraphics[width=1\linewidth]{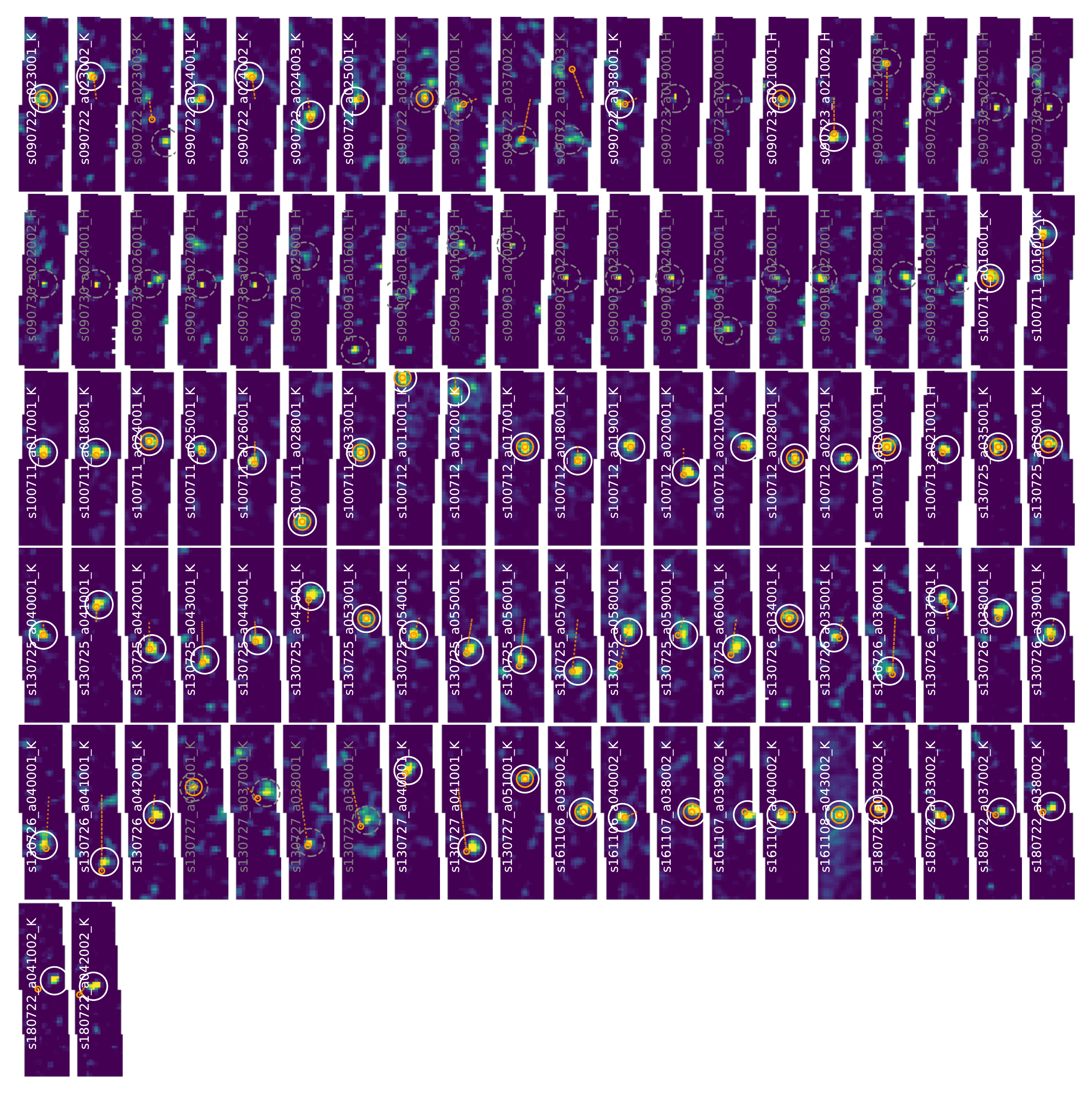}
  \caption{Individual exposure detection maps for HR 8799 b in H or K-band. }
  \label{fig:detecHR8799b}
\end{figure}

\begin{figure}
  \centering
  \includegraphics[width=1\linewidth]{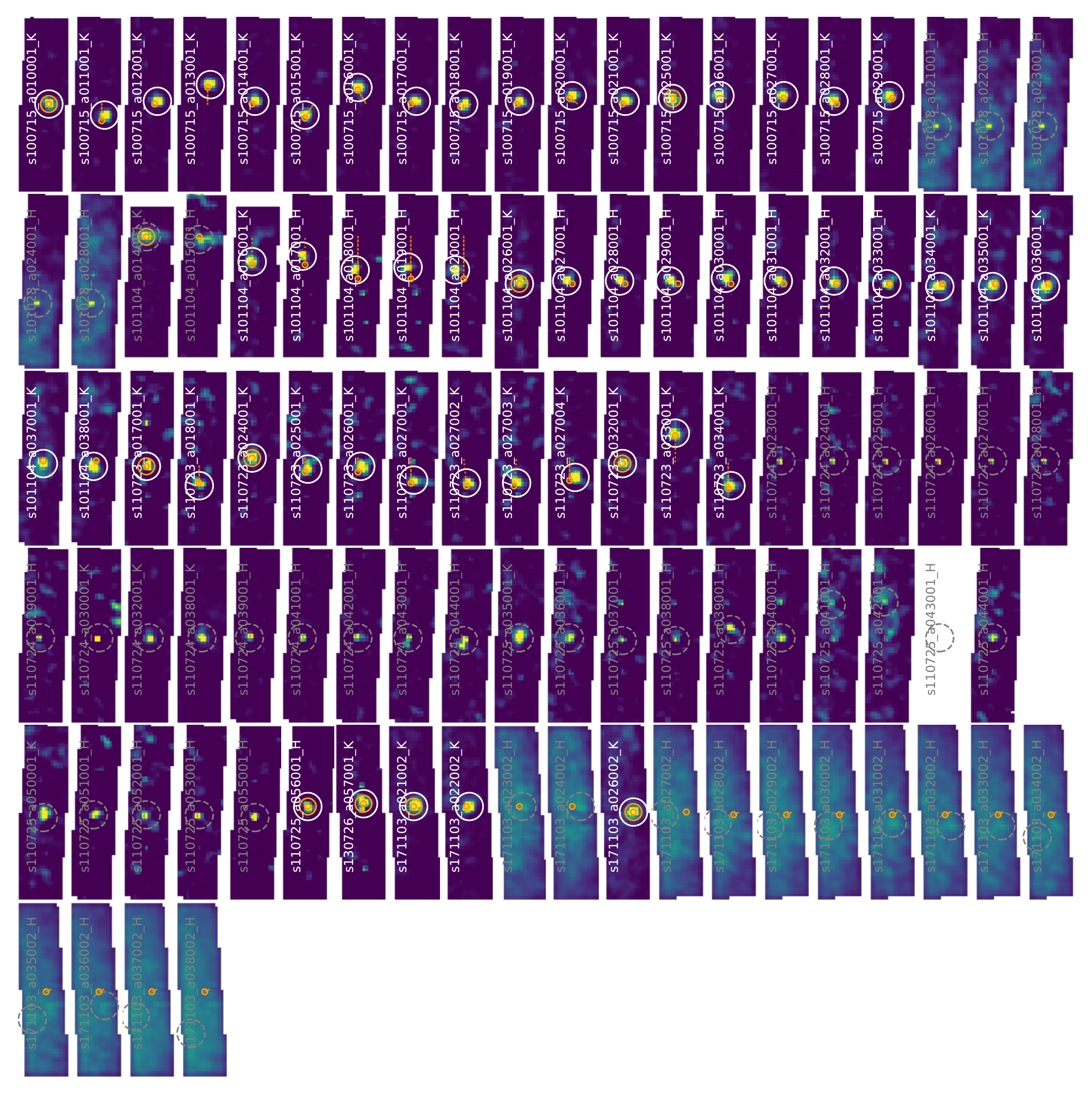}
  \caption{Individual exposure detection maps for HR 8799 c in H or K-band. }
  \label{fig:detecHR8799c}
\end{figure}

\autoref{fig:detecHR8799b} and \autoref{fig:detecHR8799c} show the individual exposure detection maps for HR 8799 b and c respectively. The corresponding filename is annotated on each frame. The white solid circles indicate the selected frames which have been used in the analyses of \Secref{sec:RV}. The greyed dashed circles represent non-detections or detections suffering from an image artifact. The pointing offsets calculated from the fits file header keywords are drawn with orange lines. The reference pointing, marked as a double orange circle, is the chosen to be the first detection in the given sequence of observations.


For the sake of completeness, we here remind the definition of both the AIC and BIC, which is based on the maximized likelihood.
\begin{equation}
    \textrm{AIC} = 2N_{\bm \phi}-2\mathrm{log}\mathcal{L}(\tilde{{\bm \phi}},\tilde{s}^2),
\end{equation}
where $N_{\bm \phi}$ the number of parameters in ${\bm \phi}$. The smaller the AIC, the more likely the model is.
\begin{equation}
    \textrm{BIC} = \mathrm{log}(N_{\bm d})N_{\bm \phi}-2\mathrm{log}\mathcal{L}(\tilde{{\bm \phi}},\tilde{s}^2),
\end{equation}

The model comparison consists in calculating the difference in the criterion for two models. In the case of AIC, we have,
\begin{equation}
    \Delta \textrm{AIC} = \textrm{AIC}_0 - \textrm{AIC}_1
\end{equation}
with the index 0 or 1 corresponding to $\mathcal{H}_0$ and $\mathcal{H}_1$ respectively.
The probability of one hypothesis compared to another is proportional to $\mathrm{exp}(\Delta \textrm{AIC}/2)$. For example, if $\Delta \textrm{AIC}=4.6$, then $\mathcal{H}_1$ is ten times more likely than $\mathcal{H}_0$. In practice, mismatch between the models and the data can make the interpretation more challenging.

\subsection{Marginalizing over linear parameters}
\label{app:nonlinesti}
We will show how to derive the posterior of the non linear parameters ${\bm \psi}$ marginalized over ${\bm \phi}$. ${\bm \phi}$ is here seen as background parameters. For example, it can be used to calculate the posterior of the radial velocity estimate of the planet, while marginalizing over the planet flux and the starlight subtraction.

The marginalized posterior of ${\bm \psi}$ is given by,
\begin{align}
    \mathcal{P}({\bm \psi} \vert {\bm d}) &= \int_{s} \int_{{\bm \phi}} \mathcal{P}({\bm \psi}, {\bm \phi}, s \vert {\bm d}),\nonumber \\
    &=  \frac{1}{\mathcal{P}({\bm d})} \int_{s} \int_{{\bm \phi}} \mathcal{P}({\bm d} \vert {\bm \psi}, {\bm \phi}, s)\mathcal{P}( {\bm \psi}, {\bm \phi}, s),\nonumber \\
    &=  \frac{1}{\mathcal{P}({\bm d})} \int_{s}  \int_{{\bm \phi}} \mathcal{P}({\bm d} \vert {\bm \psi}, {\bm \phi}, s)\mathcal{P}( {\bm \psi})\mathcal{P}( {\bm \phi})\mathcal{P}(s),\nonumber \\
    &=  \frac{\mathcal{P}( {\bm \psi})}{\mathcal{P}({\bm d})} \int_{s} \mathcal{P}(s) \int_{{\bm \phi}} \mathcal{P}({\bm d} \vert {\bm \psi}, {\bm \phi}, s)\mathcal{P}( {\bm \phi}).
\end{align}
In \autoref{eq:applikdphi}, we showed that the likelihood can be written as,
\begin{align}
    \mathcal{P}({\bm d} \vert {\bm \psi}, {\bm \phi}, s) 
    &= \frac{1}{\sqrt{(2\pi)^{N_{\bm d}}\vert{\bm \Sigma}_0\vert s^{2N_{\bm d}}}} \exp\left\lbrace-\frac{1}{2s^2}\chi^2_{0,{\bm \phi}=\tilde{{\bm \phi}},{\bm \psi}}\right\rbrace  \nonumber \\
    &* \exp\left\lbrace-\frac{1}{2s^2}\Delta{\bm \phi}^{\top}({\bm M}_{{\bm \psi}}^{\top}{\bm \Sigma}_0^{-1}{\bm M}_{{\bm \psi}})\Delta{\bm \phi}\right\rbrace.
\end{align}
Assuming unbound uniform improper priors on the elements of ${\bm \phi}$, we have,
\begin{align}
    \int_{{\bm \phi}} \mathcal{P}({\bm d} \vert {\bm \psi}, {\bm \phi}, s^2)\mathcal{P}( {\bm \phi}) &= \frac{1}{\sqrt{(2\pi)^{N_{\bm d}}\vert{\bm \Sigma}_0\vert s^{2N_{\bm d}}}} \exp\left\lbrace-\frac{1}{2s^2}\chi^2_{0,{\bm \phi}=\tilde{{\bm \phi}},{\bm \psi}}\right\rbrace \nonumber \\ &*\int_{\Delta{\bm \phi}} \exp\left\lbrace-\frac{1}{2s^2}\Delta{\bm \phi}^{\top}({\bm M}_{{\bm \psi}}^{\top}{\bm \Sigma}_0^{-1}{\bm M}_{{\bm \psi}})\Delta{\bm \phi}\right\rbrace,\nonumber \\
    &= \sqrt{\frac{(2\pi)^{N_{\bm \phi}} s^{2N_{\bm \phi}}}{(2\pi)^{N_{\bm d}} s^{2N_{\bm d}}}} 
    \frac{1}{\sqrt{\vert{\bm \Sigma}_0\vert \times \left\vert{\bm M}_{{\bm \psi}}^{\top}{\bm \Sigma}_0^{-1}{\bm M}_{{\bm \psi}}\right\vert}}
    \exp\left\lbrace-\frac{1}{2s^2}\chi^2_{0,{\bm \phi}=\tilde{{\bm \phi}},{\bm \psi}}\right\rbrace
\end{align}
Then, we can integrate over $s$, assuming improper prior $\mathcal{P}(s) \propto 1/s^\alpha$. 
\begin{align}
    \int_{s} \mathcal{P}(s) \int_{{\bm \phi}} \mathcal{P}({\bm d} \vert {\bm \psi}, {\bm \phi}, s^2)\mathcal{P}( {\bm \phi}) 
    &\propto \frac{1}{\sqrt{\vert{\bm \Sigma}_0\vert \times \left\vert{\bm M}_{{\bm \psi}}^{\top}{\bm \Sigma}_0^{-1}{\bm M}_{{\bm \psi}}\right\vert}} \nonumber \\
    &* \int_{s}  \frac{1}{s^{N_{\bm d}-N_{\bm \phi}+\alpha}}
    \exp\left\lbrace-\frac{1}{2s^2}\chi^2_{0,{\bm \phi}=\tilde{{\bm \phi}},{\bm \psi}}\right\rbrace
\end{align}
We then use the identity ($N>1$, $a>0$),
\begin{equation}
    \int_{x=0}^{\infty} \frac{1}{x^N}\exp\left\lbrace-\frac{a}{x^2}\right\rbrace = \frac{1}{2a^{(N-1)/2}}\Gamma\left[\frac{N-1}{2}\right].
\end{equation}

We derive that, 
\begin{equation}
    \int_{s} \mathcal{P}(s) \int_{{\bm \phi}} \mathcal{P}({\bm d} \vert {\bm \psi}, {\bm \phi}, s^2)\mathcal{P}( {\bm \phi}) 
    \propto \frac{1}{\sqrt{\vert{\bm \Sigma}_0\vert \times \left\vert{\bm M}_{{\bm \psi}}^{\top}{\bm \Sigma}_0^{-1}{\bm M}_{{\bm \psi}}\right\vert}} \left(\frac{1}{\chi^2_{0,{\bm \phi}=\tilde{{\bm \phi}},{\bm \psi}}}\right)^{\frac{N_{\bm d}-N_{\bm \phi}+\alpha-1}{2}},
\end{equation}
and therefore conclude that,
\begin{equation}
    \mathcal{P}({\bm \psi} \vert {\bm d})
    \propto  \frac{\mathcal{P}( {\bm \psi})}{\sqrt{\vert{\bm \Sigma}_0\vert \times \left\vert{\bm M}_{{\bm \psi}}^{\top}{\bm \Sigma}_0^{-1}{\bm M}_{{\bm \psi}}\right\vert}} \left(\frac{1}{\chi^2_{0,{\bm \phi}=\tilde{{\bm \phi}},{\bm \psi}}}\right)^{\frac{N_{\bm d}-N_{\bm \phi}+\alpha-1}{2}}.
    \label{eq:appRVpost}
\end{equation}

For numerical calculation, taking the logarithm gives the alternative expression,
\begin{align}
    \mathrm{log}\mathcal{P}({\bm \psi} \vert {\bm d})
    &= \mathrm{Cst} + \mathrm{log}\mathcal{P}( {\bm \psi}) -\frac{1}{2}\mathrm{log}\left(\vert{\bm \Sigma}_0\vert \times \left\vert{\bm M}_{{\bm \psi}}^{\top}{\bm \Sigma}_0^{-1}{\bm M}_{{\bm \psi}}\right\vert\right) \nonumber \\
    &-\frac{N_{\bm d}-N_{\bm \phi}+\alpha-1}{2} \mathrm{log}\chi^2_{0,{\bm \phi}=\tilde{{\bm \phi}},{\bm \psi}},\nonumber \\
\end{align}

Note that the minimized chi square can be written,
\begin{align}
    \chi^2_{0,{\bm \phi}=\tilde{{\bm \phi}},{\bm \psi}} &= ({\bm d}-{\bm M}\tilde{{\bm \phi}})^{\top}{\bm \Sigma}_0^{-1}({\bm d}-{\bm M}\tilde{{\bm \phi}}),\nonumber \\
    &= {\bm d}^{\top}{\bm \Sigma}_0^{-1}{\bm d} + \tilde{{\bm \phi}}^{\top}{\bm M}^{\top}{\bm \Sigma}_0^{-1}{\bm M}\tilde{{\bm \phi}} - 2{\bm d}^{\top}{\bm \Sigma}_0^{-1}{\bm M}\tilde{{\bm \phi}},\nonumber \\
    &= {\bm d}^{\top}{\bm \Sigma}_0^{-1}{\bm d} + \left[\tilde{{\bm \phi}}^{\top}{\bm M}^{\top}{\bm \Sigma}_0^{-1}{\bm M} - {\bm d}^{\top}{\bm \Sigma}_0^{-1}{\bm M}\right]\tilde{{\bm \phi}} - {\bm d}^{\top}{\bm \Sigma}_0^{-1}{\bm M}\tilde{{\bm \phi}},\nonumber \\
    &= {\bm d}^{\top}{\bm \Sigma}_0^{-1}{\bm d} - {\bm d}^{\top}{\bm \Sigma}_0^{-1}{\bm M}\tilde{{\bm \phi}},\nonumber \\
    &= {\bm d}^{\top}{\bm \Sigma}_0^{-1}{\bm d} - {\bm d}^{\top}{\bm \Sigma}_0^{-1}{\bm M}({\bm M}^{\top}{\bm \Sigma_0}^{-1}{\bm M})^{-1}{\bm M}^{\top}{\bm \Sigma_0}^{-1}{\bm d}
\end{align}
where we used the fact that the derivative of the chi square with respect to ${\bm \phi}$ vanishes, such that,
\begin{align}
    \frac{\diff \chi^2}{\diff {\bm \phi}} &= {\bm \phi}^{\top}{\bm M}^{\top}{\bm \Sigma}_0^{-1}{\bm M} - {\bm d}^{\top}{\bm \Sigma}_0^{-1}{\bm M}=0.
\end{align}

\newpage
\bibliographystyle{aasjournal}
\bibliography{bibliography}

\begin{thebibliography}{}
\expandafter\ifx\csname natexlab\endcsname\relax\def\natexlab#1{#1}\fi
\providecommand{\url}[1]{\href{#1}{#1}}

\bibitem[{{Astropy Collaboration} {et~al.}(2013){Astropy Collaboration},
  {Robitaille}, {Tollerud}, {Greenfield}, {Droettboom}, {Bray}, {Aldcroft},
  {Davis}, {Ginsburg}, {Price-Whelan}, {Kerzendorf}, {Conley}, {Crighton},
  {Barbary}, {Muna}, {Ferguson}, {Grollier}, {Parikh}, {Nair}, {Unther},
  {Deil}, {Woillez}, {Conseil}, {Kramer}, {Turner}, {Singer}, {Fox}, {Weaver},
  {Zabalza}, {Edwards}, {Azalee Bostroem}, {Burke}, {Casey}, {Crawford},
  {Dencheva}, {Ely}, {Jenness}, {Labrie}, {Lim}, {Pierfederici}, {Pontzen},
  {Ptak}, {Refsdal}, {Servillat}, \& {Streicher}}]{Astropy2013}
{Astropy Collaboration}, {Robitaille}, T.~P., {Tollerud}, E.~J., {et~al.} 2013,
  \aap, 558, A33

\bibitem[{{Baines} {et~al.}(2012){Baines}, {White}, {Huber}, {Jones},
  {Boyajian}, {McAlister}, {ten Brummelaar}, {Turner}, {Sturmann}, {Sturmann},
  {Goldfinger}, {Farrington}, {Riedel}, {Ireland }, {von Braun}, \&
  {Ridgway}}]{Baines2012}
{Baines}, E.~K., {White}, R.~J., {Huber}, D., {et~al.} 2012, \apj, 761, 57

\bibitem[{{Baraffe} {et~al.}(2003){Baraffe}, {Chabrier}, {Barman}, {Allard}, \&
  {Hauschildt}}]{Baraffe2003}
{Baraffe}, I., {Chabrier}, G., {Barman}, T.~S., {Allard}, F., \& {Hauschildt},
  P.~H. 2003, \aap, 402, 701

\bibitem[{{Barman} {et~al.}(2015){Barman}, {Konopacky}, {Macintosh}, \&
  {Marois}}]{Barman2015}
{Barman}, T.~S., {Konopacky}, Q.~M., {Macintosh}, B., \& {Marois}, C. 2015,
  \apj, 804, 61

\bibitem[{{Barman} {et~al.}(2011){Barman}, {Macintosh}, {Konopacky}, \&
  {Marois}}]{Barman2011}
{Barman}, T.~S., {Macintosh}, B., {Konopacky}, Q.~M., \& {Marois}, C. 2011,
  \apj, 733, 65

\bibitem[{{Bell} {et~al.}(2015){Bell}, {Mamajek}, \& {Naylor}}]{Bell2015}
{Bell}, C.~P.~M., {Mamajek}, E.~E., \& {Naylor}, T. 2015, \mnras, 454, 593

\bibitem[{{Birkby} {et~al.}(2013){Birkby}, {de Kok}, {Brogi}, {de Mooij},
  {Schwarz}, {Albrecht}, \& {Snellen}}]{Birkby2013}
{Birkby}, J.~L., {de Kok}, R.~J., {Brogi}, M., {et~al.} 2013, \mnras, 436, L35

\bibitem[{{Birkby} {et~al.}(2017){Birkby}, {de Kok}, {Brogi}, {Schwarz}, \&
  {Snellen}}]{Birkby2017}
{Birkby}, J.~L., {de Kok}, R.~J., {Brogi}, M., {Schwarz}, H., \& {Snellen},
  I.~A.~G. 2017, \aj, 153, 138

\bibitem[{{Blunt} {et~al.}(2019){Blunt}, {Ngo}, {Wang}, {Angelo}, {Cody}, {De
  Rosa}, {Malena}, {Rubenzahl}, \& Logan}]{blunt2019}
{Blunt}, S., {Ngo}, H., {Wang}, J., {et~al.} 2019, {sblunt/orbitize: API Update
  to Prepare for RV+Astrometry Fits}, , , doi:10.5281/zenodo.3337378.
\newblock \url{https://doi.org/10.5281/zenodo.3337378}

\bibitem[{{Boehle} {et~al.}(2016){Boehle}, {Larkin}, {Adkins}, {Aliado},
  {Fitzgerald}, {Johnson}, {Lyke}, {Magnone}, {Sohn}, {Wang}, \&
  {Weiss}}]{Boehle2016}
{Boehle}, A., {Larkin}, J.~E., {Adkins}, S.~M., {et~al.} 2016, in Society of
  Photo-Optical Instrumentation Engineers (SPIE) Conference Series, Vol. 9908,
  Ground-based and Airborne Instrumentation for Astronomy VI, 99082Q

\bibitem[{{Booth} {et~al.}(2016){Booth}, {Jord{\'a}n}, {Casassus}, {Hales},
  {Dent}, {Faramaz}, {Matr{\`a}}, {Barkats}, {Brahm}, \& {Cuadra}}]{Booth2016}
{Booth}, M., {Jord{\'a}n}, A., {Casassus}, S., {et~al.} 2016, \mnras, 460, L10

\bibitem[{{Brogi} {et~al.}(2016){Brogi}, {de Kok}, {Albrecht}, {Snellen},
  {Birkby}, \& {Schwarz}}]{Brogi2016}
{Brogi}, M., {de Kok}, R.~J., {Albrecht}, S., {et~al.} 2016, \apj, 817, 106

\bibitem[{{Brogi} {et~al.}(2014){Brogi}, {de Kok}, {Birkby}, {Schwarz}, \&
  {Snellen}}]{Brogi2014}
{Brogi}, M., {de Kok}, R.~J., {Birkby}, J.~L., {Schwarz}, H., \& {Snellen},
  I.~A.~G. 2014, \aap, 565, A124

\bibitem[{{Brogi} {et~al.}(2017){Brogi}, {Line}, {Bean}, {D{\'e}sert}, \&
  {Schwarz}}]{Brogi2017}
{Brogi}, M., {Line}, M., {Bean}, J., {D{\'e}sert}, J.~M., \& {Schwarz}, H.
  2017, \apj, 839, L2

\bibitem[{{Brogi} \& {Line}(2019)}]{Brogi2019}
{Brogi}, M., \& {Line}, M.~R. 2019, \aj, 157, 114

\bibitem[{{Brogi} {et~al.}(2012){Brogi}, {Snellen}, {de Kok}, {Albrecht},
  {Birkby}, \& {de Mooij}}]{Brogi2012}
{Brogi}, M., {Snellen}, I. A.~G., {de Kok}, R.~J., {et~al.} 2012, \nat, 486,
  502

\bibitem[{{Brogi} {et~al.}(2013){Brogi}, {Snellen}, {de Kok}, {Albrecht},
  {Birkby}, \& {de Mooij}}]{Brogi2013}
{Brogi}, M., {Snellen}, I.~A.~G., {de Kok}, R.~J., {et~al.} 2013, \apj, 767, 27

\bibitem[{{Bryan} {et~al.}(2018){Bryan}, {Benneke}, {Knutson}, {Batygin}, \&
  {Bowler}}]{Bryan2018}
{Bryan}, M.~L., {Benneke}, B., {Knutson}, H.~A., {Batygin}, K., \& {Bowler},
  B.~P. 2018, Nature Astronomy, 2, 138

\bibitem[{{Cantalloube} {et~al.}(2015){Cantalloube}, {Mouillet}, {Mugnier},
  {Milli}, {Absil}, {Gomez Gonzalez}, {Chauvin}, {Beuzit}, \&
  {Cornia}}]{Cantalloube2015}
{Cantalloube}, F., {Mouillet}, D., {Mugnier}, L.~M., {et~al.} 2015, \aap, 582,
  A89

\bibitem[{{de Kok} {et~al.}(2013){de Kok}, {Brogi}, {Snellen}, {Birkby},
  {Albrecht}, \& {de Mooij}}]{deKok2013}
{de Kok}, R.~J., {Brogi}, M., {Snellen}, I.~A.~G., {et~al.} 2013, \aap, 554,
  A82

\bibitem[{{Fabrycky} \& {Murray-Clay}(2010)}]{Fabrycky2010}
{Fabrycky}, D.~C., \& {Murray-Clay}, R.~A. 2010, \apj, 710, 1408

\bibitem[{{Follert} {et~al.}(2014){Follert}, {Dorn}, {Oliva}, {Lizon},
  {Hatzes}, {Piskunov}, {Reiners}, {Seemann}, {Stempels}, {Heiter}, {Marquart},
  {Lockhart}, {Anglada-Escude}, {L{\"o}winger}, {Baade}, {Grunhut}, {Bristow},
  {Klein}, {Jung}, {Ives}, {Kerber}, {Pozna}, {Paufique}, {Kaeufl}, {Origlia},
  {Valenti}, {Gojak}, {Hilker}, {Pasquini}, {Smette}, \&
  {Smoker}}]{Follert2014SPIE}
{Follert}, R., {Dorn}, R.~J., {Oliva}, E., {et~al.} 2014, in Society of
  Photo-Optical Instrumentation Engineers (SPIE) Conference Series, Vol. 9147,
  Ground-based and Airborne Instrumentation for Astronomy V, 914719

\bibitem[{{Foreman-Mackey} {et~al.}(2013){Foreman-Mackey}, {Hogg}, {Lang}, \&
  {Goodman}}]{ForemanMackey2013}
{Foreman-Mackey}, D., {Hogg}, D.~W., {Lang}, D., \& {Goodman}, J. 2013, \pasp,
  125, 306

\bibitem[{{Gaia Collaboration}(2018)}]{Gaia2018}
{Gaia Collaboration}. 2018, VizieR Online Data Catalog, I/345

\bibitem[{{Gontcharov}(2006)}]{Gontcharov2006}
{Gontcharov}, G.~A. 2006, Astronomy Letters, 32, 759

\bibitem[{{Gravity Collaboration} {et~al.}(2019){Gravity Collaboration},
  {Lacour}, {Nowak}, {Wang}, {Pfuhl}, {Eisenhauer}, {Abuter}, {Amorim},
  {Anugu}, {Benisty}, {Berger}, {Beust}, {Blind}, {Bonnefoy}, {Bonnet},
  {Bourget}, {Brandner}, {Buron}, {Collin}, {Charnay}, {Chapron}, {Cl{\'e}net},
  {Coud{\'e} Du Foresto}, {de Zeeuw}, {Deen}, {Dembet}, {Dexter}, {Duvert},
  {Eckart}, {F{\"o}rster Schreiber}, {F{\'e}dou}, {Garcia}, {Garcia Lopez},
  {Gao}, {Gendron}, {Genzel}, {Gillessen}, {Gordo}, {Greenbaum}, {Habibi},
  {Haubois}, {Hau{\ss}mann}, {Henning}, {Hippler}, {Horrobin}, {Hubert},
  {Jimenez Rosales}, {Jocou}, {Kendrew}, {Kervella}, {Kolb}, {Lagrange},
  {Lapeyr{\`e}re}, {Le Bouquin}, {L{\'e}na}, {Lippa}, {Lenzen}, {Maire},
  {Molli{\`e}re}, {Ott}, {Paumard}, {Perraut}, {Perrin}, {Pueyo}, {Rabien},
  {Ram{\'\i}rez}, {Rau}, {Rodr{\'\i}guez-Coira}, {Rousset}, {Sanchez-Bermudez},
  {Scheithauer}, {Schuhler}, {Straub}, {Straubmeier}, {Sturm}, {Tacconi},
  {Vincent}, {van Dishoeck}, {von Fellenberg}, {Wank}, {Waisberg}, {Widmann},
  {Wieprecht}, {Wiest}, {Wiezorrek}, {Woillez}, {Yazici}, {Ziegler}, \&
  {Zins}}]{Lacour2019}
{Gravity Collaboration}, {Lacour}, S., {Nowak}, M., {et~al.} 2019, \aap, 623,
  L11

\bibitem[{{Haffert} {et~al.}(2019){Haffert}, {Bohn}, {de Boer}, {Snellen},
  {Brinchmann}, {Girard}, {Keller}, \& {Bacon}}]{Haffert2019}
{Haffert}, S.~Y., {Bohn}, A.~J., {de Boer}, J., {et~al.} 2019, Nature
  Astronomy, 329

\bibitem[{{Hoeijmakers} {et~al.}(2018){Hoeijmakers}, {Ehrenreich}, {Heng},
  {Kitzmann}, {Grimm}, {Allart}, {Deitrick}, {Wyttenbach}, {Oreshenko}, {Pino},
  {Rimmer}, {Molinari}, \& {Di Fabrizio}}]{Hoeijmakers2018}
{Hoeijmakers}, H.~J., {Ehrenreich}, D., {Heng}, K., {et~al.} 2018, \nat, 560,
  453

\bibitem[{{Hughes} {et~al.}(2011){Hughes}, {Wilner}, {Andrews}, {Williams},
  {Su}, {Murray-Clay}, \& {Qi}}]{Hughes2011}
{Hughes}, A.~M., {Wilner}, D.~J., {Andrews}, S.~M., {et~al.} 2011, \apj, 740,
  38

\bibitem[{Hunter(2007)}]{Hunter2007}
Hunter, J.~D. 2007, Computing In Science \& Engineering, 9, 90

\bibitem[{{Husser} {et~al.}(2013){Husser}, {Wende-von Berg}, {Dreizler},
  {Homeier}, {Reiners}, {Barman}, \& {Hauschildt}}]{Husser2013}
{Husser}, T.~O., {Wende-von Berg}, S., {Dreizler}, S., {et~al.} 2013, \aap,
  553, A6

\bibitem[{{Kaeufl} {et~al.}(2004){Kaeufl}, {Ballester}, {Biereichel},
  {Delabre}, {Donaldson}, {Dorn}, {Fedrigo}, {Finger}, {Fischer}, {Franza},
  {Gojak}, {Huster}, {Jung}, {Lizon}, {Mehrgan}, {Meyer}, {Moorwood}, {Pirard},
  {Paufique}, {Pozna}, {Siebenmorgen}, {Silber}, {Stegmeier}, \&
  {Wegerer}}]{Kaeufl2004}
{Kaeufl}, H.-U., {Ballester}, P., {Biereichel}, P., {et~al.} 2004, in Society
  of Photo-Optical Instrumentation Engineers (SPIE) Conference Series, Vol.
  5492, Ground-based Instrumentation for Astronomy, ed. A.~F.~M. {Moorwood} \&
  M.~{Iye}, 1218--1227

\bibitem[{{Kanodia} \& {Wright}(2018{\natexlab{a}})}]{Kanodia2018RNAAS}
{Kanodia}, S., \& {Wright}, J. 2018{\natexlab{a}}, Research Notes of the
  American Astronomical Society, 2, 4

\bibitem[{{Kanodia} \& {Wright}(2018{\natexlab{b}})}]{Kanodia2018ascl}
{Kanodia}, S., \& {Wright}, J.~T. 2018{\natexlab{b}}, {Barycorrpy: Barycentric
  velocity calculation and leap second management}, , , ascl:1808.001

\bibitem[{{Kass} \& {Raftery}(1995)}]{Kass1995}
{Kass}, R.~E., \& {Raftery}, A.~E. 1995, Journal of the American Statistical
  Association, 90, 773.
\newblock
  \url{https://www.tandfonline.com/doi/abs/10.1080/01621459.1995.10476572}

\bibitem[{{Konopacky} {et~al.}(2013){Konopacky}, {Barman}, {Macintosh}, \&
  {Marois}}]{Konopacky2013}
{Konopacky}, Q.~M., {Barman}, T.~S., {Macintosh}, B.~A., \& {Marois}, C. 2013,
  Science, 339, 1398

\bibitem[{{Konopacky} {et~al.}(2016){Konopacky}, {Marois}, {Macintosh},
  {Galicher}, {Barman}, {Metchev}, \& {Zuckerman}}]{Konopacky2016}
{Konopacky}, Q.~M., {Marois}, C., {Macintosh}, B.~A., {et~al.} 2016, \aj, 152,
  28

\bibitem[{{Krabbe} {et~al.}(2004){Krabbe}, {Gasaway}, {Song}, {Iserlohe},
  {Weiss}, {Larkin}, {Barczys}, \& {Lafreniere}}]{Krabbe2004}
{Krabbe}, A., {Gasaway}, T., {Song}, I., {et~al.} 2004, in Society of
  Photo-Optical Instrumentation Engineers (SPIE) Conference Series, Vol. 5492,
  Ground-based Instrumentation for Astronomy, ed. A.~F.~M. {Moorwood} \&
  M.~{Iye}, 1403--1410

\bibitem[{{Larkin} {et~al.}(2006){Larkin}, {Barczys}, {Krabbe}, {Adkins},
  {Aliado}, {Amico}, {Brims}, {Campbell}, {Canfield}, {Gasaway}, {Honey},
  {Iserlohe}, {Johnson}, {Kress}, {LaFreniere}, {Magnone}, {Magnone},
  {McElwain}, {Moon}, {Quirrenbach}, {Skulason}, {Song}, {Spencer}, {Weiss}, \&
  {Wright}}]{Larkin2006}
{Larkin}, J., {Barczys}, M., {Krabbe}, A., {et~al.} 2006, New Astronomy
  Reviews, 50, 362

\bibitem[{{Lockhart} {et~al.}(2019){Lockhart}, {Do}, {Larkin}, {Boehle},
  {Campbell}, {Chappell}, {Chu}, {Ciurlo}, {Cosens}, {Fitzgerald}, {Ghez},
  {Lu}, {Lyke}, {Mieda}, {Rudy}, {Vayner}, {Walth}, \& {Wright}}]{Lockhart2019}
{Lockhart}, K.~E., {Do}, T., {Larkin}, J.~E., {et~al.} 2019, \aj, 157, 75

\bibitem[{{Lockwood} {et~al.}(2014){Lockwood}, {Johnson}, {Bender}, {Carr},
  {Barman}, {Richert}, \& {Blake}}]{Lockwood2014}
{Lockwood}, A.~C., {Johnson}, J.~A., {Bender}, C.~F., {et~al.} 2014, \apj, 783,
  L29

\bibitem[{{Lord}(1992)}]{Lord1992}
{Lord}, S.~D. 1992, {A new software tool for computing Earth's atmospheric
  transmission of near- and far-infrared radiation}, Tech. rep.

\bibitem[{{Lyke} {et~al.}(2017){Lyke}, {Do}, {Boehle}, {Campbell}, {Chappell},
  {Fitzgerald}, {Gasawy}, {Iserlohe}, {Krabbe}, {Larkin}, {Lockhard}, {Lu},
  {Mieda}, {McElwain}, {Perrin}, {Rudy}, {Sitarski}, {Vayner}, {Walth},
  {Weiss}, {Wizanski}, \& {Wright}}]{Lyke2017}
{Lyke}, J., {Do}, T., {Boehle}, A., {et~al.} 2017, {OSIRIS Toolbox:
  OH-Suppressing InfraRed Imaging Spectrograph pipeline}, , , ascl:1710.021

\bibitem[{{Marois} {et~al.}(2008){Marois}, {Macintosh}, {Barman}, {Zuckerman},
  {Song}, {Patience}, {Lafreni{\`e}re}, \& {Doyon}}]{Marois2008science}
{Marois}, C., {Macintosh}, B., {Barman}, T., {et~al.} 2008, Science, 322, 1348

\bibitem[{{Marois} {et~al.}(2010){Marois}, {Zuckerman}, {Konopacky},
  {Macintosh}, \& {Barman}}]{Marois2010}
{Marois}, C., {Zuckerman}, B., {Konopacky}, Q.~M., {Macintosh}, B., \&
  {Barman}, T. 2010, \nat, 468, 1080

\bibitem[{{Martin} {et~al.}(2014){Martin}, {Fitzgerald}, {McLean}, {Adkins},
  {Aliado}, {Brims}, {Johnson}, {Magnone}, {Wang}, \& {Weiss}}]{Martin2014SPIE}
{Martin}, E.~C., {Fitzgerald}, M.~P., {McLean}, I.~S., {et~al.} 2014, in
  Society of Photo-Optical Instrumentation Engineers (SPIE) Conference Series,
  Vol. 9147, Ground-based and Airborne Instrumentation for Astronomy V, 914781

\bibitem[{{Matthews} {et~al.}(2014){Matthews}, {Kennedy}, {Sibthorpe}, {Booth},
  {Wyatt}, {Broekhoven-Fiene}, {Macintosh}, \& {Marois}}]{Matthews2014}
{Matthews}, B., {Kennedy}, G., {Sibthorpe}, B., {et~al.} 2014, \apj, 780, 97

\bibitem[{{Mawet} {et~al.}(2017){Mawet}, {Delorme}, {Jovanovic}, {Wallace},
  {Bartos}, {Wizinowich}, {Fitzgerald}, {Lilley}, {Ruane}, {Wang}, {Klimovich},
  \& {Xin}}]{Mawet2017SPIE}
{Mawet}, D., {Delorme}, J.~R., {Jovanovic}, N., {et~al.} 2017, in Society of
  Photo-Optical Instrumentation Engineers (SPIE) Conference Series, Vol. 10400,
  1040029

\bibitem[{{Metchev} {et~al.}(2015){Metchev}, {Heinze}, {Apai}, {Flateau},
  {Radigan}, {Burgasser}, {Marley}, {Artigau}, {Plavchan}, \&
  {Goldman}}]{Metchev2015}
{Metchev}, S.~A., {Heinze}, A., {Apai}, D., {et~al.} 2015, \apj, 799, 154

\bibitem[{{Mieda} {et~al.}(2014){Mieda}, {Wright}, {Larkin}, {Graham},
  {Adkins}, {Lyke}, {Campbell}, {Maire}, {Do}, \& {Gordon}}]{Mieda2014}
{Mieda}, E., {Wright}, S.~A., {Larkin}, J.~E., {et~al.} 2014, \pasp, 126, 250

\bibitem[{{Nugroho} {et~al.}(2017){Nugroho}, {Kawahara}, {Masuda}, {Hirano},
  {Kotani}, \& {Tajitsu}}]{Nugroho2017}
{Nugroho}, S.~K., {Kawahara}, H., {Masuda}, K., {et~al.} 2017, \aj, 154, 221

\bibitem[{{O'Neil} {et~al.}(2019){O'Neil}, {Martinez}, {Hees}, {Ghez}, {Do},
  {Witzel}, {Konopacky}, {Becklin}, {Chu}, {Lu}, {Matthews}, \&
  {Sakai}}]{ONeil2019}
{O'Neil}, K.~K., {Martinez}, G.~D., {Hees}, A., {et~al.} 2019, \aj, 158, 4

\bibitem[{{Petit dit de la Roche} {et~al.}(2018){Petit dit de la Roche},
  {Hoeijmakers}, \& {Snellen}}]{PetitditdelaRoche2018}
{Petit dit de la Roche}, D.~J.~M., {Hoeijmakers}, H.~J., \& {Snellen}, I.~A.~G.
  2018, \aap, 616, A146

\bibitem[{{Pueyo} {et~al.}(2015){Pueyo}, {Soummer}, {Hoffmann}, {Oppenheimer},
  {Graham}, {Zimmerman}, {Zhai}, {Wallace}, {Vescelus}, {Veicht}, {Vasisht},
  {Truong}, {Sivaramakrishnan}, {Shao}, {Roberts}, {Roberts}, {Rice}, {Parry},
  {Nilsson}, {Lockhart}, {Ligon}, {King}, {Hinkley}, {Hillenbrand}, {Hale},
  {Dekany}, {Crepp}, {Cady}, {Burruss}, {Brenner}, {Beichman}, \&
  {Baranec}}]{Pueyo2015}
{Pueyo}, L., {Soummer}, R., {Hoffmann}, J., {et~al.} 2015, \apj, 803, 31

\bibitem[{{Reidemeister} {et~al.}(2009){Reidemeister}, {Krivov}, {Schmidt},
  {Fiedler}, {M{\"u}ller}, {L{\"o}hne}, \& {Neuh{\"a}user}}]{Reidemeister2009}
{Reidemeister}, M., {Krivov}, A.~V., {Schmidt}, T.~O.~B., {et~al.} 2009, \aap,
  503, 247

\bibitem[{{Rodler} {et~al.}(2012){Rodler}, {Lopez-Morales}, \&
  {Ribas}}]{Rodler2012}
{Rodler}, F., {Lopez-Morales}, M., \& {Ribas}, I. 2012, \apjl, 753, L25

\bibitem[{{Ruffio} {et~al.}(2017){Ruffio}, {Macintosh}, {Wang}, {Pueyo},
  {Nielsen}, {De Rosa}, {Czekala}, {Marley}, {Arriaga}, {Bailey}, {Barman},
  {Bulger}, {Chilcote}, {Cotten}, {Doyon}, {Duch{\^e}ne}, {Fitzgerald},
  {Follette}, {Gerard}, {Goodsell}, {Graham}, {Greenbaum}, {Hibon}, {Hung},
  {Ingraham}, {Kalas}, {Konopacky}, {Larkin}, {Maire}, {Marchis}, {Marois},
  {Metchev}, {Millar-Blanchaer}, {Morzinski}, {Oppenheimer}, {Palmer},
  {Patience}, {Perrin}, {Poyneer}, {Rajan}, {Rameau}, {Rantakyr{\"o}},
  {Savransky}, {Schneider}, {Sivaramakrishnan}, {Song}, {Soummer}, {Thomas},
  {Wallace}, {Ward-Duong}, {Wiktorowicz}, \& {Wolff}}]{Ruffio2017}
{Ruffio}, J.-B., {Macintosh}, B., {Wang}, J.~J., {et~al.} 2017, \apj, 842, 14

\bibitem[{{Salz} {et~al.}(2018){Salz}, {Czesla}, {Schneider}, {Nagel},
  {Schmitt}, {Nortmann}, {Alonso-Floriano}, {L{\'o}pez-Puertas}, {Lamp{\'o}n},
  {Bauer}, {Snellen}, {Pall{\'e}}, {Caballero}, {Yan}, {Chen}, {Sanz-Forcada},
  {Amado}, {Quirrenbach}, {Ribas}, {Reiners}, {B{\'e}jar}, {Casasayas-Barris},
  {Cort{\'e}s-Contreras}, {Dreizler}, {Guenther}, {Henning}, {Jeffers},
  {Kaminski}, {K{\"u}rster}, {Lafarga}, {Lara}, {Molaverdikhani}, {Montes},
  {Morales}, {S{\'a}nchez-L{\'o}pez}, {Seifert}, {Zapatero Osorio}, \&
  {Zechmeister}}]{Salz2018}
{Salz}, M., {Czesla}, S., {Schneider}, P.~C., {et~al.} 2018, \aap, 620, A97

\bibitem[{{Schwarz} {et~al.}(2015){Schwarz}, {Brogi}, {de Kok}, {Birkby}, \&
  {Snellen}}]{Schwarz2015}
{Schwarz}, H., {Brogi}, M., {de Kok}, R., {Birkby}, J., \& {Snellen}, I. 2015,
  \aap, 576, A111

\bibitem[{{Schwarz} {et~al.}(2016){Schwarz}, {Ginski}, {de Kok}, {Snellen},
  {Brogi}, \& {Birkby}}]{Schwarz2016}
{Schwarz}, H., {Ginski}, C., {de Kok}, R.~J., {et~al.} 2016, \aap, 593, A74

\bibitem[{{Snellen} {et~al.}(2014){Snellen}, {Brandl}, {de Kok}, {Brogi},
  {Birkby}, \& {Schwarz}}]{Snellen2014}
{Snellen}, I. A.~G., {Brandl}, B.~R., {de Kok}, R.~J., {et~al.} 2014, \nat,
  509, 63

\bibitem[{{Snellen} {et~al.}(2010){Snellen}, {de Kok}, {de Mooij}, \&
  {Albrecht}}]{Snellen2010}
{Snellen}, I. A.~G., {de Kok}, R.~J., {de Mooij}, E. J.~W., \& {Albrecht}, S.
  2010, \nat, 465, 1049

\bibitem[{{Soummer} {et~al.}(2011){Soummer}, {Hagan}, {Pueyo}, {Thormann},
  {Rajan}, \& {Marois}}]{Soummer2011}
{Soummer}, R., {Hagan}, J.~B., {Pueyo}, L., {et~al.} 2011, \apj, 741, 55

\bibitem[{{Su} {et~al.}(2009){Su}, {Rieke}, {Stapelfeldt}, {Malhotra},
  {Bryden}, {Smith}, {Misselt}, {Moro-Martin}, \& {Williams}}]{Su2009}
{Su}, K.~Y.~L., {Rieke}, G.~H., {Stapelfeldt}, K.~R., {et~al.} 2009, \apj, 705,
  314

\bibitem[{{Teachey} \& {Kipping}(2018)}]{Teachey2018}
{Teachey}, A., \& {Kipping}, D.~M. 2018, Science Advances, 4, eaav1784

\bibitem[{{Vanderburg} {et~al.}(2018){Vanderburg}, {Rappaport}, \&
  {Mayo}}]{Vanderburg2018}
{Vanderburg}, A., {Rappaport}, S.~A., \& {Mayo}, A.~W. 2018, \aj, 156, 184

\bibitem[{{Vousden} {et~al.}(2016){Vousden}, {Farr}, \& {Mandel}}]{Vousden2016}
{Vousden}, W.~D., {Farr}, W.~M., \& {Mandel}, I. 2016, \mnras, 455, 1919

\bibitem[{{Wang} {et~al.}(2018{\natexlab{a}}){Wang}, {Mawet}, {Fortney},
  {Hood}, {Morley}, \& {Benneke}}]{WangJi2018}
{Wang}, J., {Mawet}, D., {Fortney}, J.~J., {et~al.} 2018{\natexlab{a}}, \aj,
  156, 272

\bibitem[{{Wang} {et~al.}(2018{\natexlab{b}}){Wang}, {Graham}, {Dawson},
  {Fabrycky}, {De Rosa}, {Pueyo}, {Konopacky}, {Macintosh}, {Marois}, {Chiang},
  {Ammons}, {Arriaga}, {Bailey}, {Barman}, {Bulger}, {Chilcote}, {Cotten},
  {Doyon}, {Duch{\^e}ne}, {Esposito}, {Fitzgerald}, {Follette}, {Gerard},
  {Goodsell}, {Greenbaum}, {Hibon}, {Hung}, {Ingraham}, {Kalas}, {Larkin},
  {Maire}, {Marchis}, {Marley}, {Metchev}, {Millar-Blanchaer}, {Nielsen},
  {Oppenheimer}, {Palmer}, {Patience}, {Perrin}, {Poyneer}, {Rajan}, {Rameau},
  {Rantakyr{\"o}}, {Ruffio}, {Savransky}, {Schneider}, {Sivaramakrishnan},
  {Song}, {Soummer}, {Thomas}, {Wallace}, {Ward-Duong}, {Wiktorowicz}, \&
  {Wolff}}]{Wang2018}
{Wang}, J.~J., {Graham}, J.~R., {Dawson}, R., {et~al.} 2018{\natexlab{b}}, \aj,
  156, 192

\bibitem[{{Wertz} {et~al.}(2017){Wertz}, {Absil}, {G{\'o}mez Gonz{\'a}lez},
  {Milli}, {Girard}, {Mawet}, \& {Pueyo}}]{Wertz2017}
{Wertz}, O., {Absil}, O., {G{\'o}mez Gonz{\'a}lez}, C.~A., {et~al.} 2017, \aap,
  598, A83

\bibitem[{{Wilner} {et~al.}(2018){Wilner}, {MacGregor}, {Andrews}, {Hughes},
  {Matthews}, \& {Su}}]{Wilner2018}
{Wilner}, D.~J., {MacGregor}, M.~A., {Andrews}, S.~M., {et~al.} 2018, The
  Astrophysical Journal, 855, 56

\bibitem[{{Zucker}(2003)}]{Zucker2003}
{Zucker}, S. 2003, \mnras, 342, 1291

\bibitem[{{Zuckerman} {et~al.}(2011){Zuckerman}, {Rhee}, {Song}, \&
  {Bessell}}]{Zuckerman2011}
{Zuckerman}, B., {Rhee}, J.~H., {Song}, I., \& {Bessell}, M.~S. 2011, \apj,
  732, 61

\bibitem[{{Zurlo} {et~al.}(2016){Zurlo}, {Vigan}, {Galicher}, {Maire}, {Mesa},
  {Gratton}, {Chauvin}, {Kasper}, {Moutou}, {Bonnefoy}, {Desidera}, {Abe},
  {Apai}, {Baruffolo}, {Baudoz}, {Baudrand}, {Beuzit}, {Blancard},
  {Boccaletti}, {Cantalloube}, {Carle}, {Cascone}, {Charton}, {Claudi},
  {Costille}, {de Caprio}, {Dohlen}, {Dominik}, {Fantinel}, {Feautrier},
  {Feldt}, {Fusco}, {Gigan}, {Girard}, {Gisler}, {Gluck}, {Gry}, {Henning},
  {Hugot}, {Janson}, {Jaquet}, {Lagrange}, {Langlois}, {Llored}, {Madec},
  {Magnard}, {Martinez}, {Maurel}, {Mawet}, {Meyer}, {Milli},
  {Moeller-Nilsson}, {Mouillet}, {Orign{\'e}}, {Pavlov}, {Petit}, {Puget},
  {Quanz}, {Rabou}, {Ramos}, {Rousset}, {Roux}, {Salasnich}, {Salter},
  {Sauvage}, {Schmid}, {Soenke}, {Stadler}, {Suarez}, {Turatto}, {Udry},
  {Vakili}, {Wahhaj}, {Wildi}, \& {Antichi}}]{Zurlo2016}
{Zurlo}, A., {Vigan}, A., {Galicher}, R., {et~al.} 2016, \aap, 587, A57

\end{thebibliography}

\end{document}